\documentclass[aps,prd,reprint,twocolumn,superscriptaddress,longbibliography,nofootinbib,floatfix,showpacs]{revtex4-1}
\usepackage{epsfig}
\usepackage{amsmath,amssymb,amsfonts}
\usepackage{hyperref}
\usepackage{mathrsfs}
\usepackage{bbm}
\usepackage{slashed}
\usepackage{graphicx}
\usepackage{verbatim}

\usepackage{bm}

\usepackage[usenames]{xcolor}

\definecolor{darkgreen}{rgb}{0.2,0.6,0}

\newcommand{\be}{\begin{equation}}
\newcommand{\ee}{\end{equation}}
\newcommand{\bw}{\begin{widetext}}
\newcommand{\ew}{\end{widetext}}
\newcommand{\bi}{\begin{itemize}}
\newcommand{\ei}{\end{itemize}}
\newcommand{\ud}{\mathrm{d}}

\newcommand{\LCperp}{{\scriptscriptstyle \perp}}

\usepackage[T1]{fontenc} \usepackage[latin1]{inputenc}

\begin{document}

\title{Worldline instantons for the momentum spectrum of Schwinger pair production in space-time dependent fields}

\author{Gianluca Degli Esposti}
\email{g.degli-esposti@hzdr.de}
\affiliation{Helmholtz-Zentrum Dresden-Rossendorf, Bautzner Landstra{\ss}e 400, 01328 Dresden, Germany}

\author{Greger Torgrimsson}
\email{greger.torgrimsson@umu.se}
\affiliation{Department of Physics, Ume{\aa} University, SE-901 87 Ume{\aa}, Sweden}

\begin{abstract}

We show how to use the worldline-instanton formalism to calculate the momentum spectrum of the electron-positron pairs produced by an electric field that depends on both space and time. Using the LSZ reduction formula with a worldline representation for the propagator in a spacetime field, we make use of the saddle-point method to obtain a semiclassical approximation of the pair-production spectrum. In order to check the final result, we integrate the spectrum and compare with the results obtained using a previous instanton method for the imaginary part of the effective action.

\end{abstract}
\maketitle
\section{Introduction}

The creation of particle pairs in the presence of a strong electric field is a theoretical prediction of Quantum Electrodynamics (QED) which was first suggested by Sauter \cite{Sauter:1931}.
Schwinger \cite{Schwinger:1951} calculated the rate of pair production at one loop level in a constant electric field. Today, this process is also commonly referred to as Schwinger pair production. 
The key feature is that the exponential scaling with respect to the electric field strength\footnote{We use units with $c=\hbar=m_e=1$ and absorb $eE\to E$.}
\be\label{eq:constant}
\mathbb P \sim e^{-\frac{\pi}{E}}
\ee
implies that this is a \textit{nonperturbative} effect, as \eqref{eq:constant} does not have an expansion in powers of $E$ when $E \to 0$. Due to the exponential scaling with respect to the electric field strength, and the critical field being of the order of $10^{18}$ V/m, to this day it has not been possible to observe this effect experimentally. 

There are very few exactly solvable field shapes, see e.g.~\cite{Ilderton:2014mla,Breev:2021lpn},  
so one has to either turn to a full numerical treatment or to approximate methods in order to deal with non-constant fields. However, it is challenging to obtain numerical results for spacetime dependent fields\footnote{Progress has been made, though, in the last couple of years, and~\cite{Aleksandrov:2017mtq,Kohlfurst:2022vwf} have now shown how to numerically obtain the spectrum for fields that depend on $t$ and $z$ and with components in the $x-y$ plane, with one important example being the field of two head-on colliding plane waves.} and for the physically relevant regime $E\ll1$. On the other hand, for $E\ll1$ we can instead turn to semiclassical approximations. Time-dependent fields have been studied in \cite{Brezin:1970,Popov:1972,Popov:2005,Dunne:2005wi,Dunne:2006fp}. The semiclassical approximation can in principle be obtained by directly solving the Dirac equation using a WKB approach. However, while this has been done for time dependent fields, there is so far no WKB method that can be used for general spacetime dependent fields; see~\cite{Kohlfurst:2021skr} for the most recent study. A more promising approach is the worldline instanton method~\cite{Affleck:1981bma,Dunne:2005wi,Dunne:2006fp,Dunne:2006ur,Dumlu:2011cc}. It is at this point a well developed method for obtaining the total/integrated probability from the imaginary part of the effective action. Indeed, a numerical code based on discretized instantons\footnote{Discretized instantons were also used in~\cite{Gould:2017fve} for pair production by a combination of a constant electric field and a thermal background.} was developed in~\cite{Torgrimsson:2017cyb,Schneider:2018huk} that can be used for general fields. As an example, in~\cite{Schneider:2018huk} it was applied to the e-dipole field~\cite{Gonoskov:2012}, which is an exact solution to Maxwell's equations that is localized in all four spacetime coordinates. The only other method that has been able to deal with such a complicated field~\cite{Gonoskov:2013ada} is the locally-constant-field approximation (LCFA), where one takes the constant-field result and replaces the volume factor with a spacetime integral, $V^4 \, \mathbb{P} (F_{\mu\nu})\to\int\ud^4x \, \mathbb{P}(F(x))$. For fields below the Schwinger limit, one can perform the $\ud^4x$ integral with the saddle-point method, and then the result agrees with the slowly-varying-field limit of the instanton approximation. Thus, the instanton approximation contains physics beyond the LCFA\footnote{If one does not perform that $\ud^4x$ integral in the LCFA with the saddle-point method, then in principle the LCFA contains physics beyond the weak-field limit, which would be relevant if one approaches the Schwinger limit.} but can still be used for realistic 4D fields.   
Our main goal is therefore to develop the instanton method so that we can also obtain the momentum spectrum and not just the total probability.   

Starting from the LSZ reduction formula, the key feature is the representation of the exact propagator as a particle path integral \cite{Feynman:1948,Feynman:1950,Feynman:1951gn} with an integration over proper time \cite{Schwinger:1951}. In a previous paper \cite{DegliEsposti:2021its} we showed how to use the worldline instanton method to compute, on the amplitude level, the spectrum of pair production in a time-dependent background field. Unlike the method used in \cite{Dunne:2005wi,Dunne:2006fp,Dunne:2006ur,Torgrimsson:2017cyb,Schneider:2018huk}, which uses periodic (closed) instantons $x^\mu(0) = x^\mu(1)$ with the topology of the circle, our instantons have open boundary conditions and describe free particles asymptotically. 
Thus, our instantons are complex near the tunneling region but approach real particle trajectories at asymptotic times.
Schwinger pair production by a constant electric field has been studied with open worldlines in~\cite{Barut:1989mc,Rajeev:2021zae}. 

For a time-dependent field it was possible to obtain simple analytic expressions for the full amplitude using saddle-point approximations for the worldline integrals. In particular, the final prefactor is trivial because all the contributions cancel each other out. Not surprisingly, this is no longer true for more general fields. In the present work we focus on linearly polarized fields which depend on time and one space variable $A(z,t)$ such that $E(z,t) \to 0$ asymptotically. In particular, the $z$ dependence breaks the translation symmetry along the $z$ axis, therefore the momentum along said direction is no longer conserved. In practice, as we shall see, this means that $p_3 + p^\prime_3 \neq 0$ is allowed, where $p$ and $p^\prime$ are respectively the electron and positron momenta. We will show that the particle momenta are related to the asymptotic conditions of the instantons.

In the ``closed-loop'' method, the starting point is the effective action in the worldline representation~\cite{Strassler:1992zr,Schubert:2001}, i.e. in terms of a path integral over all \textit{closed} spacetime trajectories. Using unitarity, one obtains the total probability of pair production. The main advantage is that the initial expression is simpler and there is no need to worry about external states or amputations. However, this method only gives the probability summed over all possible external momenta. The method we show here provides the momentum spectrum, and performing all the momentum integrals with the saddle-point method gives agreement with the closed-loop method.

We use the saddle-point approximation in order to compute the integrals analytically. However, we find the saddle points of the path integral numerically, also known as ``instantons'', which are solution to the Lorentz force equation with asymptotic conditions determined by the momenta. 

Although the solutions themselves depend on the parametrization, the final probability does not. We show that one can for example choose a simple parametrization tilted in the complex plane or a nontrivial one, such that the instantons are parallel to the real axis asymptotically. In Sec.~\ref{sec:Defs} we give some basic definitions. In Sec.~\ref{sec:GY} we apply the Gelfand-Yaglom method to compute the path integral prefactor. 
In Sec.~\ref{sec:OrdInt} we calculate the ordinary integrals over the spacetime coordinates and proper time. It is eventually possible to obtain a simple expression of this contribution in the asymptotic time limit. Sec.~\ref{sec:Spin} we calculate the spin sum, putting everything together in Sec.~\ref{sec:FinalFormula}.
We explain the procedure to find the instantons numerically in Sec.~\ref{sec:Instantons}.
Once we have all the contributions to the spectrum, we make use once again of the saddle-point method to determine the widths of the spectrum, in addition to the integrated/total probability, 
in Sec.~\ref{sec:PerpIntegral} and \ref{sec:LongIntegral}. We also show the agreement with the discrete instanton method for the effective action used in~\cite{Schneider:2018huk,Torgrimsson:2017cyb}. Finally, in Sec.~\ref{sec:kToZero} we consider the limit where the spatial dependence becomes very slow compared to the time dependence, which gives a regularized volume factor rather than the infinite volume factor one finds if one starts with a purely time-dependent field. 
This is done by expanding the instantons in a suitable way.

\section{Basic definitions} \label{sec:Defs}

Our starting point is the worldline representation~\cite{Feynman:1951gn} of the dressed fermion propagator\footnote{Different representations can be found in~\cite{Fradkin:1991ci,Gies:2005ke,Ahmadiniaz:2020wlm,Corradini:2020prz}.} in a background field $A_\mu$,
\begin{widetext}
\be\label{propagatorWorldline}
S(x,x')=(i\slashed{\mathcal{D}}_x+m)\frac{1}{2}\int_0^\infty\!\ud T\int\limits_{q(0)=x'}^{q(1)=x}\mathcal{D}q\exp\left\{-i\left[\frac{Tm^2}{2}+\int_0^1\!\ud\tau\left(\frac{\dot{q}^2}{2T}+A(q)\dot{q}\right)\right]\right\}\mathcal{P}\exp\left\{-i\frac{T}{4}\int_0^1\ud\tau\sigma^{\mu\nu}F_{\mu\nu}\right\} \;,
\ee
where $\mathcal{D}_\mu=\partial_\mu+iA_\mu$, $F_{\mu\nu}=\partial_\mu A_\nu-\partial_\nu A_\mu$, $\mathcal{P}$ means proper-time ordering and $\sigma^{\mu\nu}=\frac{i}{2}[\gamma^\mu,\gamma^\nu]$. Note that~\eqref{propagatorWorldline} holds for an arbitrary electromagnetic field. We obtain the amplitude by amputating using the LSZ reduction formula
\be\label{LSZ3pair}
M=\lim_{t\to+\infty}\lim_{t'\to+\infty}\int\ud^3x\ud^3 x'\,e^{ip_jx^j}\bar{u}_r^{(\rm asymp)}(t,{\bf p})\gamma^0S(x,x')\gamma^0 e^{ip'_jx'^j}v_{r'}^{(\rm asymp)}(t',{\bf p}') \;,
\ee
\end{widetext}
where $p_j x^j=\sum_{i=1}^3p_j x^j$, and the asymptotic states can be replaced by their WKB/adiabatic approximations,
\be\label{UandV}
\begin{split}
	{\sf U}_r(t,{\bf q})&=(\gamma^0\pi_0+\gamma^i\pi_i+1){\sf G}^+(t,{\bf q})R_r \\
	{\sf V}_r(t,-{\bf q})&=(-\gamma^0\pi_0+\gamma^i\pi_i+1){\sf G}^-(t,{\bf q})R_r \;,
\end{split}
\ee
where ($t_r\in\mathbb{R}$ is a constant)
\be\label{Gpmdef}
{\sf G}^\pm(t,{\bf q})=[2\pi_0(\pi_0\pm\pi_3)]^{-\frac{1}{2}}\exp\bigg[\mp i\int_{t_r}^t\!\ud t'\,\pi_0(t')\bigg] \;,
\ee
the spin basis is chosen according to $\gamma^0\gamma^3 R_s=R_s$, $r=1,2$, and
$\pi_\LCperp=q_\LCperp$, $\pi_3(t)=q_3-A(t)$, $\pi_0=\sqrt{m_\LCperp^2+\pi_3^2(t)}$, $m_\LCperp=\sqrt{1+q_\LCperp^2}$. (Recall, we work with units where $m=1$.) These nontrivial solutions are in general needed for fields with $A(t=-\infty)\ne A(t=+\infty)$. However, in this paper we will focus on fields that also depend on $z$, such as
\be\label{SauterPulseA3}
A_3=\frac{E}{\omega}\tanh(\omega t)\text{sech}^2(k z) \;,
\ee
and, although $A(t=-\infty)\ne A(t=+\infty)$, the worldline starts and ends at $|z|\to\infty$, so we have $A_3\to0$ asymptotically. So, for any nonzero $k$ we have a rather different case compared to if one starts with $k=0$. 
We will refer to the field in~\eqref{SauterPulseA3} as ``the Sauter pulse'', since $E(t,z)=E\text{sech}^2(\omega t)\text{sech}^2(kz)$ is the product of a Sauter pulse in $t$ and a Sauter pulse in $z$.

\section{Path integral and Gelfand-Yaglom}\label{sec:GY}

We start by performing the path integral. Expanding around the instanton (i.e. the saddle point) gives us a quadratic path integral that we can perform using the Gelfand-Yaglom method~\cite{Dunne:2006fp}. We change variables and notation as
\be
q_\mu(\tau)\to q_\mu(\tau)+\delta q_\mu(\tau) \;,
\ee
so that from now on $q_\mu(\tau)$ is the instanton and $\delta q_\mu(\tau)$ is the integration variable. The exponent in~\eqref{propagatorWorldline} with $\sigma^{\mu\nu}$ does not influence the instanton equations and we can take $\delta q\to0$ there, since $TF_{\mu\nu}\sim\mathcal{O}(E^0)$ is not large (the other exponential terms scale like $1/E$, which is the justification for using the saddle-point method). The instantons are therefore determined by the Lorentz-force equation,
\be
\ddot{q}^\mu=TF^{\mu\nu}\dot{q}_\nu \;.
\ee 
For the class of fields that we consider in this paper we have
\be\label{LorentzTE}
\ddot{t}=TE(t,z)\dot{z}
\qquad
\ddot{z}=TE(t,z)\dot{t} \;,
\ee
where $E(t,z)=\partial A_3(t,z)/\partial t$. It will often be convenient to change proper-time variable from $\tau$ to $u=T(\tau-\sigma)$, where $\sigma$ is a constant such that the instanton passes through the field around $u\sim0$. For a symmetric instanton it would be convenient to choose $\sigma=1/2$, but, in general, different choices of $\sigma$ just corresponds to arbitrary shifts of the $u$ variable.

The $\delta q$ integrand is given by
\be\label{pathQuadLambda}
\exp\left\{-\frac{i}{2T}\int_0^1\begin{pmatrix}
	\delta t & \delta z 
\end{pmatrix}
\Lambda
\begin{pmatrix}
	\delta t \\ \delta z
\end{pmatrix}
\right\} \;,
\ee
where
\be
\Lambda=T\begin{pmatrix} -\frac{1}{T}\partial_\tau^2+A_{tt}\dot{z} & A_{tz}\dot{z}+A_t\partial_\tau \\ A_{tz}\dot{z}-\partial_\tau A_t & \frac{1}{T}\partial_\tau^2-A_{tz}\dot{t} \end{pmatrix} \;,
\ee
where $A_{tz}=\partial^2 A_3/\partial t\partial z$ etc. and $\partial_\tau A_t=A_{tt}\dot{t}+A_{tz}\dot{z}+A_t\partial_\tau$. The path integral can be performed using the Gelfand-Yaglom method, which gives 
\be\label{detLambdaphiphi}
\det\Lambda=(\phi^{(1)}_1\phi^{(2)}_2-\phi^{(1)}_2\phi^{(2)}_1)|_{\tau=1} \;,
\ee
in terms of two solutions, $\phi^{(1)}$ and $\phi^{(2)}$, of the Jacobi equation \cite{Dunne:2006fp}
\be\label{LambdaEq}
\Lambda\phi=0 
\ee
with initial conditions
\be
\phi^{(1)}(0)=\phi^{(2)}(0)=\begin{pmatrix} 0\\0\end{pmatrix}
\ee
and
\be\label{initialdphi}
\dot{\phi}^{(1)}(0)=\begin{pmatrix} 1\\0\end{pmatrix}
\qquad
\dot{\phi}^{(2)}(0)=\begin{pmatrix} 0\\1\end{pmatrix} \;.
\ee
\eqref{LambdaEq} can also be expressed as
%\be
%\left[-\sigma_3\partial_u^2+E(t,z)i\sigma_2\partial_u+\{z'(u),-t'(u)\}\nabla E(t,z)\right]\phi=0 \;,
%\ee
\be
\left[-\partial_u^2+E(t,z)\sigma_1\partial_u+\{z'(u),t'(u)\}\nabla E(t,z)\right]\phi=0 \;,
\ee
where $\sigma_1$ is one of the Pauli matrices and
\be
\nabla E=\{A_{tt},A_{tz}\} \;.
\ee
For the normalization we use the free part ($\mathcal{D}\delta q=\mathcal{D}\delta q_0\mathcal{D}\delta q_1\mathcal{D}\delta q_2\mathcal{D}\delta q_3$)
\be\label{freePathInt}
\int_{\delta q(0)=0}^{\delta q(1)=0}\mathcal{D}\delta q\exp\left\{-i\int_0^1\ud\tau\frac{\delta\dot{q}^2}{2T}\right\} =\frac{1}{(2\pi T)^2} \;.
\ee
So the nontrivial part is given by $\sqrt{\det\Lambda_{\rm free}}/\sqrt{\det\Lambda}=1/\sqrt{\det\Lambda}$.

Our instantons have nontrivial behavior in the region where $|E|$ is significantly nonzero and straight lines in the asymptotic regions, i.e. outside the field. While~\eqref{LambdaEq} can be solved numerically as it stands, it is better to separate out the asymptotic parts. This will allow us to show analytically that $t_0$ and $t_1$ drop out in the $t_0,t_1\to\infty$ limit. 

Before the instanton enters the field region we have (``$\approx$'' could be replaced by ``$=$'' for a field with finite support)
\be
\phi^{(1)}\approx\begin{pmatrix} \tau \\ 0 \end{pmatrix}
\qquad
\phi^{(2)}\approx\begin{pmatrix} 0 \\ \tau \end{pmatrix} \;.
\ee
We follow this solution to a point $\tilde{u}_0$ where $E(t(\tilde{u}_0),z(\tilde{u}_0))$ starts to be significantly nonzero, where
\be
\tau=\frac{u-u_0}{T} \;.
\ee 
At this point we start the numerical computation with initial conditions 
\be
\phi^{(1)}(\tilde{u}_0)\approx\begin{pmatrix} (\tilde{u}_0-u_0)/T \\ 0 \end{pmatrix}
\qquad
\frac{\ud\phi^{(1)}}{\ud u}(\tilde{u}_0)\approx\begin{pmatrix} 1/T \\ 0 \end{pmatrix}
\ee
and similarly for $\phi^{(2)}$. For $t_0\to\infty$ we have ($t'=\ud t/\ud u$)
\be\label{u0u0}
\tilde{u}_0-u_0=\int_{t_0}^{\tilde{t}_0}\frac{\ud t}{t'}=\frac{t_0}{p'_0}+\mathcal{O}(1) \;,
\ee
where we have anticipated that $t'\to-p'_0$ at $u_0$. In general, we can already at this point anticipate the values of the saddle points for all integration variables and substitute them into $\Lambda$ and $T$ in~\eqref{pathQuadLambda}. We will show below that the saddle-point value of $T$ is, in the asymptotic limit, given by
\be\label{Tfromt0t1}
T=\frac{t_0}{p'_0}+\frac{t_1}{p_0}+\mathcal{O}(1) \;,
\ee
so $T$ is on the same order of magnitude as $t_0,t_1$. Thus
\be
\phi^{(1)}(\tilde{u}_0)\approx\begin{pmatrix} t_0/(Tp'_0) \\ 0 \end{pmatrix}+\mathcal{O}(1/T) \;.
\ee
At this point it might thus seem like we could approximate $\frac{\ud\phi^{(1)}}{\ud u}(\tilde{u}_0)\approx\{0,0\}$, but we will show that we need to keep this derivative equal to $\{1/T,0\}$. Instead we write
\be
\phi^{(j)}=\frac{t_0}{Tp'_0}\phi_d^{(j)}+\frac{1}{T}\phi_n^{(j)} \;,
\ee 
where
\be
\phi_d^{(1)}(\tilde{u}_0)=\begin{pmatrix} 1 \\ 0 \end{pmatrix}
\qquad
\frac{\ud\phi_d^{(1)}}{\ud u}(\tilde{u}_0)=\begin{pmatrix} 0 \\ 0 \end{pmatrix} \;,
\ee
\be
\phi_n^{(1)}(\tilde{u}_0)=\begin{pmatrix} 0 \\ 0 \end{pmatrix}
\qquad
\frac{\ud\phi_n^{(1)}}{\ud u}(\tilde{u}_0)=\begin{pmatrix} 1 \\ 0 \end{pmatrix}
\ee
and similarly for $\phi_d^{(2)}$ and $\phi_n^{(2)}$. 

We always have one solution to $\Lambda\phi=0$ given by $\phi=\{t',z'\}$. Since $\{t'',z''\}=0$ outside the field, we can write
\be
\begin{split}
\phi_d^{(1)}(u)&=c_1\{t'(u),z'(u)\}-c_2\varphi(u) \\
\phi_d^{(2)}(u)&=c_2\{t'(u),z'(u)\}+c_1\varphi(u) \;,
\end{split}
\ee
where the two constants are given by
\be
c_1=\frac{t'(\tilde{u}_0)}{t^{\prime2}(\tilde{u}_0)+z^{\prime2}(\tilde{u}_0)}
\qquad
c_2=\frac{z'(\tilde{u}_0)}{t^{\prime2}(\tilde{u}_0)+z^{\prime2}(\tilde{u}_0)}
\ee
and where $\varphi$ is a second independent solution, with orthogonal initial conditions
\be
\varphi(\tilde{u}_0)=\begin{pmatrix} -z'(\tilde{u}_0) \\ t'(\tilde{u}_0) \end{pmatrix}
\qquad
\frac{\ud\varphi}{\ud u}(\tilde{u}_0)=\begin{pmatrix} 0 \\ 0 \end{pmatrix} \;.
\ee
The contribution to~\eqref{detLambdaphiphi} involving products of $\phi_d^{(j)}$ and no $\phi_n^{(j)}$ can now be expressed as
\be
\begin{split}
D_1(u):&=\left(\frac{t_0}{Tp'_0}\right)^2(\phi_{d1}^{(1)}\phi_{d2}^{(2)}-\phi_{d2}^{(1)}\phi_{d1}^{(2)})\\
&=\left(\frac{t_0}{Tp'_0}\right)^2\frac{t'(u)\varphi_2(u)-z'(u)\varphi_1(u)}{t^{\prime2}(\tilde{u}_0)+z^{\prime2}(\tilde{u}_0)}
\end{split} \;.
\ee 
In the end we only need the asymptotic value of $D_1(u_1)$ ($u_1$ corresponds to $\tau=1$). Since $\varphi$ grows linearly in $u$, we have
\be
D_1(u_1)\approx(u_1-\tilde{u}_1)D_1'(\tilde{u}_1) \;,
\ee
where $\tilde{u}_1$ is chosen such that, for $u>\tilde{u}_1$, the instanton has left the region where the field is significantly nonzero. In the asymptotic limit we have (cf.~\eqref{u0u0})
\be
u_1-\tilde{u}_1=\frac{t_1}{p_0}+\mathcal{O}(1/T) \;.
\ee
We therefore have
\be
D_1(u_1)\approx\frac{t_0t_1}{Tp'_0p_0}\frac{t_0}{Tp'_0}h(\tilde{u}_1) \;,
\ee
where
\be\label{hfromvarphi}
h(u)=\frac{t'(u)\varphi_2'(u)-z'(u)\varphi_1'(u)}{t^{\prime2}(\tilde{u}_0)+z^{\prime2}(\tilde{u}_0)} \;.
\ee

A second contribution is given by cross terms between $\phi_d^{(j)}$ and $\phi_n^{(j)}$,
\be
D_2:=\frac{t_0}{T^2p'_0}(\phi_{d1}^{(1)}\phi_{n2}^{(2)}+\phi_{n1}^{(1)}\phi_{d2}^{(2)}-\phi_{d2}^{(1)}\phi_{n1}^{(2)}-\phi_{n2}^{(1)}\phi_{d1}^{(2)}) \;.
\ee
In the asymptotic limit, $D_2(u)$ grows quadratically in $u$, so we have  
\be
D_2(u_1)\approx\frac{1}{2}(u_1-\tilde{u}_1)^2D_2''(\tilde{u}_1) \;.
\ee
Taking the derivatives and throwing away terms with $\phi''$, $t''$ or $z''$, which anyway vanish asymptotically, we find
\be
D_2(u_1)\approx\frac{t_0t_1}{Tp'_0p_0}\frac{t_1}{Tp_0}g(\tilde{u}_1) \;,
\ee
where
\be
\begin{split}
g(u)=&c_1(\phi_{n1}^{(1)\prime}(u)\varphi_2'(u)-\phi_{n2}^{(1)\prime}(u)\varphi_1'(u))\\
+&c_2(\phi_{n1}^{(2)\prime}(u)\varphi_2'(u)-\phi_{n2}^{(2)\prime}(u)\varphi_1'(u)) \;.
\end{split}
\ee

$D_1(u)$ and $D_2(u)$ were introduced as functions of $u$ (which in the end we only need to evaluate at $u_1$). Now we have introduced two other functions of $u$, $h(u)$ and $g(u)$. The reason for doing this is that $h(\tilde{u}_1)=g(\tilde{u}_1)$ and we can prove this by showing that we in fact also have $h(u)=g(u)$, which we in turn can show by considering $h'(u)$ and $g'(u)$. In contrast, $D_1(u)$ and $D_2(u)$ are not proportional, they have different behavior at finite $u$.

To simplify $h'(u)$ and $g'(u)$, we first note that 
\be
\frac{\ud}{\ud u}(t'\phi'_1-z'\phi'_2)=0
\ee   
for any solution to $\Lambda\phi=0$. This gives us a constant of motion 
\be\label{nuDefinition}
\nu(\phi)=t'\phi'_1-z'\phi'_2 \;.
\ee
Using the initial conditions at $\tilde{u}_0$ we find
$\nu(\varphi)=0$, $\nu(\phi^{(1)})=t'(\tilde{u}_0)$ and $\nu(\phi^{(2)})=-z'(\tilde{u}_0)$.
Second, we also note that for any two solutions, $\phi$ and $\tilde{\phi}$, we have
\be
\frac{\ud}{\ud u}(\phi'_1\tilde{\phi}'_2-\phi_2'\tilde{\phi}'_1)=\nabla E\cdot(\nu(\phi)\tilde{\phi}-\nu(\tilde{\phi})\phi) 
\ee
and
\be\label{nablaEphi}
\frac{\ud}{\ud u}(t'\phi'_2-z'\phi'_1)=m_\LCperp^2\nabla E\cdot\phi \;.
\ee
Using these relations we find
\be
h'(u)=\frac{m_\LCperp^2}{t^{\prime2}(\tilde{u}_0)+z^{\prime2}(\tilde{u}_0)}\nabla E\cdot\varphi=g'(u) \;.
\ee
Since $h(\tilde{u}_0)=g(\tilde{u}_0)=0$, we therefore have $h(u)=g(u)$.

The third contribution,
\be
D_3=\frac{1}{T^2}(\phi_{n1}^{(1)}\phi_{n2}^{(2)}-\phi_{n2}^{(1)}\phi_{n1}^{(2)}) \;,
\ee
is negligible, because, asymptotically, $D_3(u)$ grows quadratically in $u$, which makes $D_3\sim\mathcal{O}(1)$, while $D_1\sim D_2\sim\mathcal{O}(T)$. 

Thus, together with~\eqref{Tfromt0t1} we find
\be\label{detLambdaFromh}
\det\Lambda\approx D_1+D_2\approx\frac{t_0t_1}{Tp'_0p_0}h(\tilde{u}_1) \;.
\ee
Note that these approximate signs become exact in the asymptotic limit $t_0,t_1\to\infty$, which we will always take in the end.
Remarkably, we see from~\eqref{hfromvarphi} that we only need to find one solution to $\Lambda\phi=0$, namely $\phi=\varphi$. 

We can simplify further by writing
\be
\hat{\varphi}(u)=\frac{\varphi(u)}{t'^2(\tilde{u}_0)+z'^2(\tilde{u}_0)}
\ee
and then we note that any solution can be expressed as
\be\label{etaDefinition}
\phi(u)=\{t'(u),z'(u)\}\chi(u)+\{-z'(u),t'(u)\}\frac{\eta(u)}{t^{\prime2}(u)+z^{\prime2}(u)} \;,
\ee
where we thus use $\chi$ and $\eta$ rather than $\phi_1$ and $\phi_2$ to represent the two degrees of freedom. From~\eqref{nuDefinition} we find
\be
m_\LCperp^2\chi'=\nu+E\frac{m_\LCperp^4-4t^{\prime2}z^{\prime2}}{(t^{\prime2}+z^{\prime2})^2}\eta+\frac{2t'z'}{t^{\prime2}+z^{\prime2}}\eta' \;,
\ee
which allows us to write~\eqref{nablaEphi} solely in terms of $\eta$,
\be\label{etaEq}
\eta''-(E^2+\nabla E\cdot\{z',t'\})\eta+\nu E=0 \;.
\ee 
This is useful because (cf.~\eqref{hfromvarphi})
\be
\{-z',t'\}\cdot\phi'=\eta'+\left(m_\LCperp^2\chi-\frac{2t'z'}{t^{\prime2}+z^{\prime2}}\eta\right)E \;,
\ee
so in the asymptotic limit this quantity also only involves $\eta$,
\be
\{-z',t'\}\cdot\phi\to\eta' \;.
\ee
Thus, using this in~\eqref{hfromvarphi} we find
\be\label{hFrometa}
h(\tilde{u}_1)=\eta'(\tilde{u}_1) \;,
\ee
where $\eta$ is the solution of~\eqref{etaEq} with $\nu=0$ and
\be\label{etaInitialForh}
\eta(\tilde{u}_0)=1 
\qquad
\eta'(\tilde{u}_0)=0 \;.
\ee

To summarize, we started with an expression for $\det\Lambda$ in terms of two two-component solutions, $\phi^{(1)}$ and $\phi^{(2)}$; separated out the factors of $t_0,t_1$ by expressing $\det\Lambda$ in terms of three two-component solutions, $\varphi$, $\phi_n^{(1)}$ and $\phi_n^{(2)}$; then showed that only $\varphi$ is needed; and finally we have now showed that we only need to find one one-component solution, $\eta$ from~\eqref{etaEq} with initial conditions as in~\eqref{etaInitialForh}. Thus, the calculation of the functional determinant~\eqref{detLambdaFromh} is greatly simplified.

\section{Ordinary integrals}\label{sec:OrdInt}

We begin with the perpendicular integrals, which are trivial relative to the other integrals. We make a shift in the path integration variable, 
\be
q_\LCperp(\tau)\to x'_\LCperp+\tau(x_\LCperp-x'_\LCperp)+q_\LCperp(\tau) \;,
\ee
so that the new $q_\LCperp(\tau)$ has boundary conditions $q_\LCperp(0)=q_\LCperp(1)=0$. Then
\be
\frac{i}{2T}\int_0^1\ud\tau\dot{q}_\LCperp^2\to\frac{i}{2T}(x_\LCperp-x'_\LCperp)^2+\frac{i}{2T}\int_0^1\ud\tau\dot{q}_\LCperp^2 
\ee
and the $q_\LCperp$ integral is just the free one (which gives one factor of $1/2\pi T$ in~\eqref{freePathInt}). With $x'_\LCperp=\varphi_\LCperp-\theta_\LCperp/2$ and $x_\LCperp=\varphi_\LCperp+\theta_\LCperp/2$, the $\varphi_\LCperp$ integral gives
\be\label{deltaPerp}
(2\pi)^2\delta^2(p_\LCperp+p'_\LCperp)
\ee
and the Gaussian $\theta_\LCperp$ integral gives a factor of 
\be\label{thetaPerpInt}
2\pi T 
\ee
to the prefactor. 

Next we turn to the integrals over $z_0=z(\tau=0)$, $z_1=z(\tau=1)$ and $T$, which we will also perform using the saddle-point method. We obtain the saddle-point equations by differentiating the exponent in~\eqref{propagatorWorldline}, where $q_\mu$ is now the instanton solution. At this point in the calculation, $q_\mu$ depends on $z_0$, $z_1$ and $T$. If we make a variation in $z_0$, $z_1$ or $T$ then that leads to a variation in $q_\mu$, which we denote $\delta'q$\footnote{We used $\delta q$ for the path integration variable.}. The corresponding variation of the exponent is given by   
\be\label{curlyBracket}
\begin{split}
&\delta'i\left\{p_3z_1+p'_3z_0-\frac{Tm_\LCperp^2}{2}-\int_0^1\!\ud\tau\left(\frac{\dot{q}^2}{2T}+A(q)\dot{q}\right)\right\}= \\
&-\frac{i}{2}(m_\LCperp^2-a^2)\delta'T+ip_3\delta'z_1+ip'_3\delta'z_0-i\Big(\frac{\dot{q}}{T}+A\Big)\delta'q\Big|_0^1 \;,
\end{split}
\ee
where
\be\label{aDefinition}
a^2=\frac{\dot{t}^2-\dot{z}^2}{T^2}=t^{\prime2}-z^{\prime2}
\ee
is a constant of motion. Thus, with $\{...\}$ denoting the curly brackets in~\eqref{curlyBracket},
\be\label{z0saddleEq}
\frac{\partial}{\partial z_0}i\{...\}=i[p'_3-z'(u_0)]
\ee
\be\label{z1saddleEq}
\frac{\partial}{\partial z_1}i\{...\}=i[p_3+z'(u_1)]
\ee
and
\be\label{TsaddleEq}
\frac{\partial}{\partial T}i\{...\}=\frac{i}{2}(a^2-m_\LCperp^2) \;.
\ee
The saddle points, $z_0^s(p'_3,p_3,m_\LCperp)$ etc., are determined by setting~\eqref{z0saddleEq}, \eqref{z1saddleEq} and~\eqref{TsaddleEq} to zero.
These equations might look rather complicated since they involve the instanton solution, which we have not found yet and which we in general can only find numerically. Fortunately, these equations simplify considerably in the asymptotic limit $t_0,t_1\to\infty$. We have 
\be\label{z0asymptoticEq}
z_0=\tilde{z}_0+\int_{\tilde{t}_0}^{t_0}\ud t\frac{z'}{t'}=-\frac{t_0z'(u_0)}{\sqrt{a^2+z^{\prime2}(u_0)}}+\mathcal{O}(1) \;,
\ee
\be\label{z1asymptoticEq}
z_1=\tilde{z}_1+\int_{\tilde{t}_1}^{t_1}\ud t\frac{z'}{t'}=\frac{t_1z'(u_1)}{\sqrt{a^2+z^{\prime2}(u_1)}}+\mathcal{O}(1) 
\ee
and
\be\label{TasymptoticEq}
T=\int_{t_0}^{t_1}\frac{\ud t}{t'}=\frac{t_1}{\sqrt{a^2+z^{\prime2}(u_1)}}+\frac{t_0}{\sqrt{a^2+z^{\prime2}(u_0)}}+\mathcal{O}(1) \;,
\ee
where we have used $t'(u_1)=\sqrt{a^2+z^{\prime2}(u_1)}$ and $t'(u_0)=-\sqrt{a^2+z^{\prime2}(u_1)}$, which follows from~\eqref{aDefinition}. We can now solve~\eqref{z0asymptoticEq}, \eqref{z1asymptoticEq} and~\eqref{TasymptoticEq},
\be
z'(u_0)=-\frac{z_0}{T}\left(1+\frac{\sqrt{t_1^2-z_1^2}}{\sqrt{t_0^2-z_0^2}}\right) \;,
\ee
\be
z'(u_1)=\frac{z_1}{T}\left(1+\frac{\sqrt{t_0^2-z_0^2}}{\sqrt{t_1^2-z_1^2}}\right)
\ee
and
\be
a^2=\frac{1}{T^2}\left(\sqrt{t_0^2-z_0^2}+\sqrt{t_1^2-z_1^2}\right)^2 \;.
\ee
By substituting these into~\eqref{z0saddleEq}, \eqref{z1saddleEq} and~\eqref{TsaddleEq} we find
\be
z_0^s=-\frac{p'_3t_0}{\sqrt{m_\LCperp^2+p_3^{\prime2}}}
\qquad
z_1^s=-\frac{p_3t_1}{\sqrt{m_\LCperp^2+p_3^2}}
\ee
and
\be
T_s=\frac{t_0}{\sqrt{m_\LCperp^2+p_3^{\prime2}}}+\frac{t_1}{\sqrt{m_\LCperp^2+p_3^2}} \;.
\ee
With $p'_0=\sqrt{m_\LCperp^2+p_3^{\prime2}}$ and $p_0=\sqrt{m_\LCperp^2+p_3^2}$ we have thus proven~\eqref{Tfromt0t1}, which we anticipated in the calculation of the functional determinant.
We will show below that the dominant contribution comes from $p'_3\sim-p_3$, so $z_0^s$ and $z_1^s$ have different signs, i.e. the instanton starts and ends on opposite sides of the field, or, in other words, the electron and positron end up at opposite sides.

It is now also straightforward to calculate the Hessian matrix ${\bf H}$ by differentiating~\eqref{z0saddleEq}, \eqref{z1saddleEq} and~\eqref{TsaddleEq}. This gives us a Gaussian integral
\be
\int\ud^3{\bf X}\exp\left\{-{\bf X}\cdot{\bf H}\cdot{\bf X}\right\} \;,
\ee
where ${\bf X}=\{\delta z_0,\delta z_1,\delta T\}$ with $\delta z_0=z_0-z_0^s$ etc. The expression for ${\bf H}$ is not particularly illuminating, but its determinant is given by
\be\label{detH}
\det{\bf H}=\frac{ip_0^{\prime3}p_0^3}{8m_\LCperp^2t_0t_1T} \;.
\ee 
Thus, the contribution from the ordinary integrals is actually quite simple. We can now see that the factors of $t_0,t_1$ cancel when combining~\eqref{detLambdaFromh} and~\eqref{detH}.

\section{Spin part}\label{sec:Spin}

Now we turn to the spin part of the prefactor. Using
\be
\frac{\ud}{\ud u}\ln[t'(u)\pm z'(u)]=\pm E(t,z) 
\ee
we can actually perform the propertime integral analytically,
\be
-\frac{iT}{4}\sigma^{\mu\nu}\int_0^1\!\ud\tau\, F_{\mu\nu}=\frac{1}{2}\gamma^0\gamma^3\ln\rho \;,
\ee
where
\be
\rho=\frac{t'(u_1)+z'(u_1)}{t'(u_0)+z'(u_0)}=-\frac{p_0-p_3}{p'_0-p'_3} \;.
\ee
Note that this does not depend on the field. For the $(i\slashed{\mathcal{D}}_x+1)$ part in~\eqref{propagatorWorldline} we can perform partial integration in $x$ and in the asymptotic limit we find 
\be
(i\slashed{\mathcal{D}}_x+1)\to\slashed{p}+1 \;.
\ee
So the calculation of the spin part reduces to
\be
\begin{split}
S=&\bar{R}_s(\slashed{p}+1)\gamma^0(\slashed{p}+1)\frac{1}{2\sqrt{\rho}}\left(\rho[1+\gamma^0\gamma^3]+1-\gamma^0\gamma^3\right)\\
&\times\gamma^0(-\slashed{p}'+1)R_{s'} \;,
\end{split}
\ee
which is field independent. At this point $S$ does perhaps not look symmetric in $p\leftrightarrow p'$, but after some algebra we find that the amplitude is proportional to $\delta_{ss'}$, so summing over spins gives a trivial factor of $2$, and in the end we find
\be\label{finalSpinPart}
\sum_{ss'}|S|^2=8p_0p'_0 \;.
\ee  

So, for the fields we consider here, the spin factor is rather simple. But when applying these methods to more general spacetime dependent fields, e.g. with magnetic components, one could study both the momentum and the spin dependence of the probability.

\section{Final formula for the general case}\label{sec:FinalFormula}

Combining all the separate contributions we finally find
\be\label{finalFormula}
\begin{split}
\mathbb P=&\int\frac{\ud^3p}{(2\pi)^3}\frac{\ud^3p'}{(2\pi)^3}\eqref{finalSpinPart}\bigg|\frac{1}{2}\eqref{freePathInt}\eqref{deltaPerp}\eqref{thetaPerpInt}\frac{\pi^{3/2}}{\sqrt{\eqref{detH}}}\frac{e^{\dots}}{\sqrt{\eqref{detLambdaFromh}}}\bigg|^2\\
=&V_\LCperp\int\frac{\ud^2p_\LCperp}{(2\pi)^2}\frac{\ud p_3}{2\pi}\frac{\ud p'_3}{2\pi}\frac{4\pi m_\LCperp^2}{p_0p'_0|h(\tilde{u}_1)|}e^{-\mathcal{A}} \;.
\end{split}
\ee
where $V_\LCperp=V_1V_2$ is a perpendicular volume factor, and
\be\label{generalFinExp}
\mathcal{A}=-2\text{Re }i\int_{-\infty}^\infty\!\ud u\; q^\mu\partial_\mu A_\nu\frac{\ud q^\nu}{\ud u} \;.
\ee
The factors of $t_0$ and $t_1$ have canceled and we can now take $t_0,t_1\to\infty$. 

\section{Instantons and the choice of einbein}\label{sec:Instantons}

We have calculated~\eqref{finalFormula} without actually having to find any instantons. However, in contrast to the purely time-dependent case~\cite{DegliEsposti:2021its}, to evaluate~\eqref{finalFormula} for a spacetime dependent field we do need to find the instanton. When writing ``the'' instanton, one should keep in mind the point made in~\cite{DegliEsposti:2021its}, i.e. that different choices of complex einbeins lead to different trajectories. Two instantons that can be (continuously) deformed to each other by deforming the einbein are equivalent and give the same results, but some choices of einbeins might be more convenient to work with and others might facilitate a physical interpretation.  

Before turning to einbeins, we first note that when expressed in terms of $E$, $\gamma_\omega = \frac{\omega}{E}$ and $\gamma_k = \frac{k}{E}$, the instantons do not have any nontrivial dependence on $E$. This follows from the fact that $E\ll1$ is the expansion parameter (i.e. one should not expect to see any complicated functions of $E$) and the final result has a simple dependence on $E$,
\be\label{EscalingP}
\mathbb{P}=E^a\mathcal{F}(\gamma_\omega,\gamma_k)\exp\left\{-\frac{\mathcal{G}(\gamma_\omega,\gamma_k)}{E}\right\} \;,
\ee
where $a$ is a constant. 
To remove the trivial $E$ dependence from the instantons we first write the instanton equations~\eqref{LorentzTE}
\be\label{Lorentzu}
t''=E(t,z)z'
\qquad
z''=E(t,z)t' \;,
\ee
where the field can be expressed as
\be
E(t,z)=E F(\omega t,kz) \;.
\ee
From this we see that we should rescale
\be\label{rescaleqwithE}
q\to\frac{q}{E}
\ee
so that the arguments become $F(\gamma_\omega t,\gamma_k z)$. To remove the remaining $E$ from~\eqref{Lorentzu} we have to rescale
\be\label{rescaleuwithE}
u\to\frac{u}{E} \;.
\ee
The rescaled instanton equations are thus given by
\be\label{LorentzRe}
t''=F(\gamma_\omega t,\gamma_k z)z'
\qquad
z''=F(\gamma_\omega t,\gamma_k z)t' \;.
\ee
From~\eqref{generalFinExp} we also see that this gives an exponent on the form~\eqref{EscalingP}. 
For the prefactor, the only nontrivial contribution in~\eqref{finalFormula} comes from $h$. From~\eqref{etaEq} with $\nu=0$ and initial conditions~\eqref{etaInitialForh} we see that $\eta$ is not rescaled with $E$, so from~\eqref{hFrometa} we see that $h=E\mathcal{H}(\gamma_\omega,\gamma_k)$. Thus, the prefactor in~\eqref{EscalingP} scales as $1/E$, i.e. $a=-1$, for the momentum spectrum. As we will show below, each of the four momentum integrals can be performed with the saddle-point method, so each of them gives a factor of $\sqrt{E}$ to the prefactor. Thus, the prefactor for the total/integrated probability scales as $E$, i.e. $a=1$. 

\begin{figure}[!ht]
  \centering
\includegraphics[width=\linewidth]{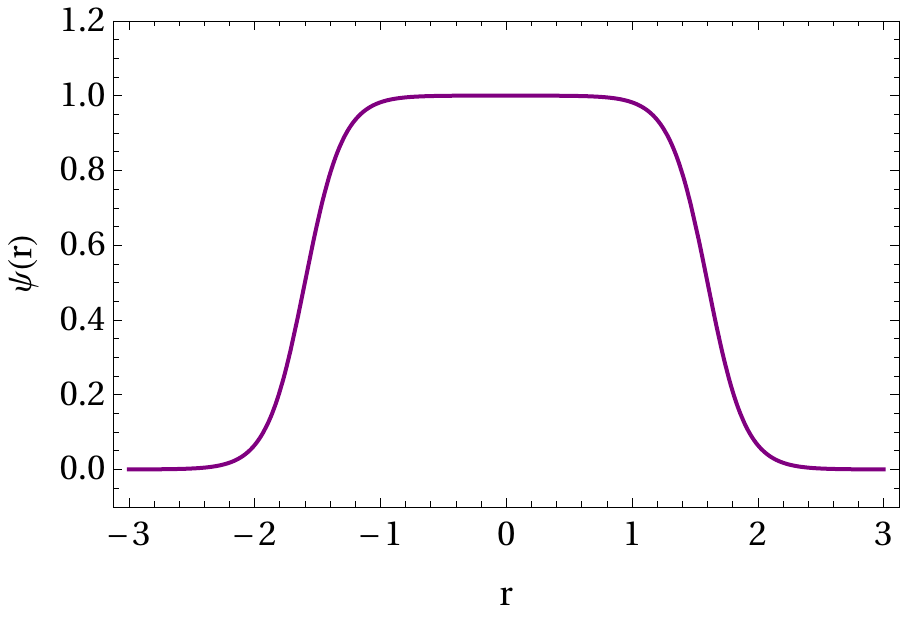}
\vspace{-.4cm}
\caption{$\psi(r)$ for parameters $L = 1.6$ and $W = 0.3$.}
\label{fig:einbein}
\end{figure}

We now turn to the choice of einbein.
We choose a nontrivial parametrization
\be
\frac{du}{dr} = f(r)
\ee
such that $f(0) = e^{-i \theta}$ and $f(r) \to 1$ asymptotically by simply defining a weighted combination
\be\label{eq:Einbein}
f(r) = 1 + (e^{-i \theta} - 1) \psi(r)
\ee
where $\psi(r)$ is some normalized bump function with $\psi(r) \sim 1$ when $r \sim 0$ and $\psi(r) \to 0$ when $r$ becomes large. As an example, we can take
\be
\psi(r) = \frac{1}{2}\left( \tanh \left[ \frac{r + L}{W} \right] + \tanh \left[ \frac{-r + L}{W} \right] \right) \; ,
\ee
which is illustrated in Fig.~\ref{fig:einbein}.
The parameters $L$ and $W$ can be chosen such that the bump becomes larger and steeper at will. 

\begin{figure*}[!ht]
\centering
\includegraphics[width=.32\linewidth]{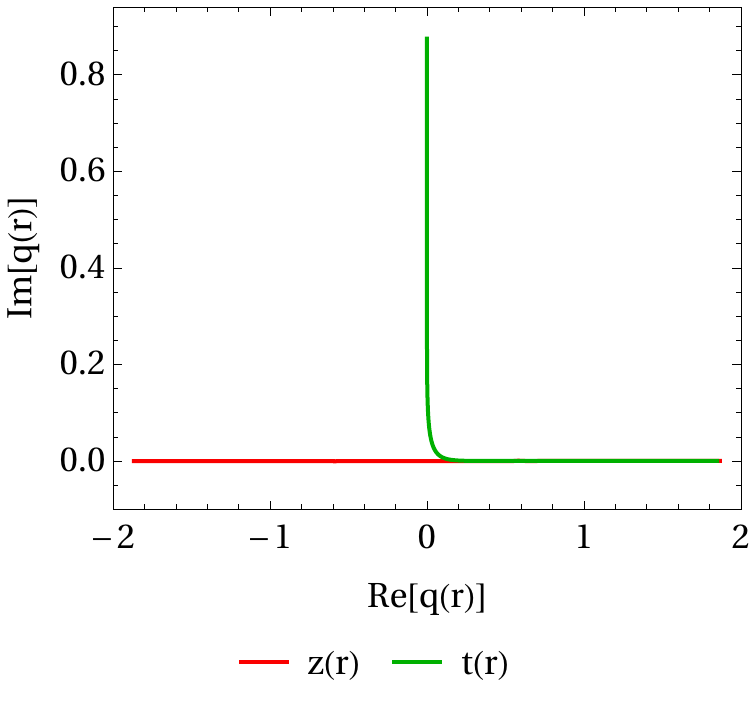}
\includegraphics[width=.32\linewidth]{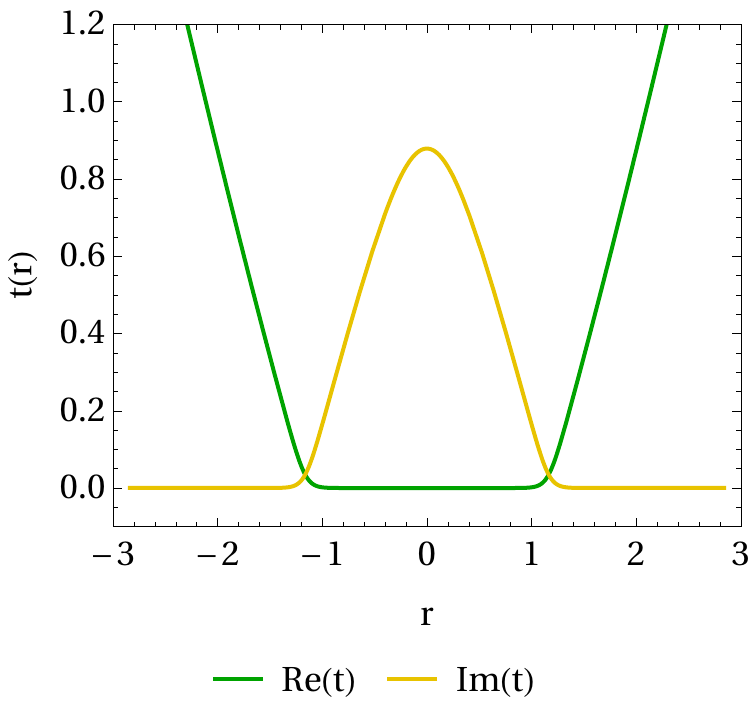}
\includegraphics[width=.32\linewidth]{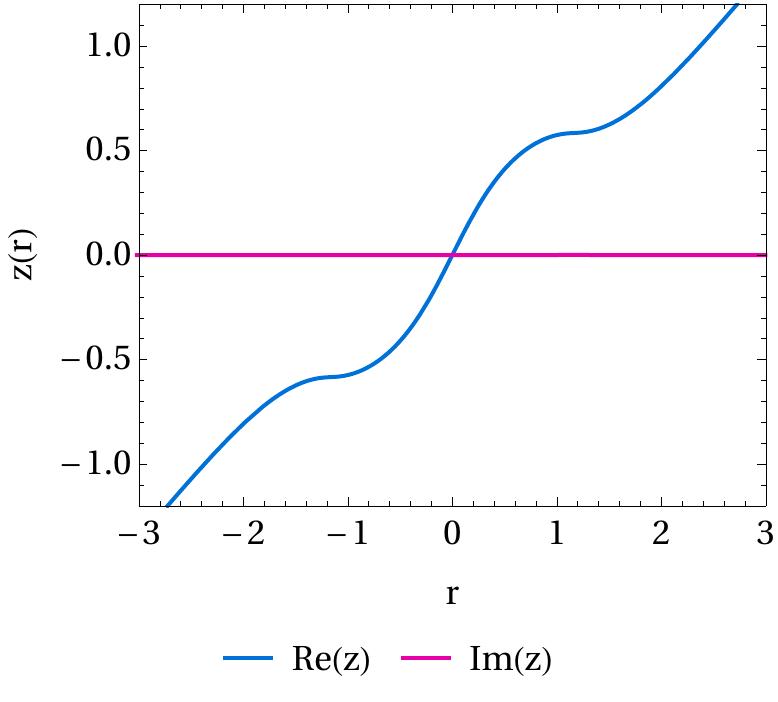}
\caption{Real and imaginary parts of $t$ and $z$ of the physical instantons, with $W = 0.1$ and $L = 1.1636$. The field parameters are $\gamma_k = \gamma_\omega = 1$.}
\label{fig:instreal}
\end{figure*}

\begin{figure}[!ht]
  \centering
\includegraphics[width=\linewidth]{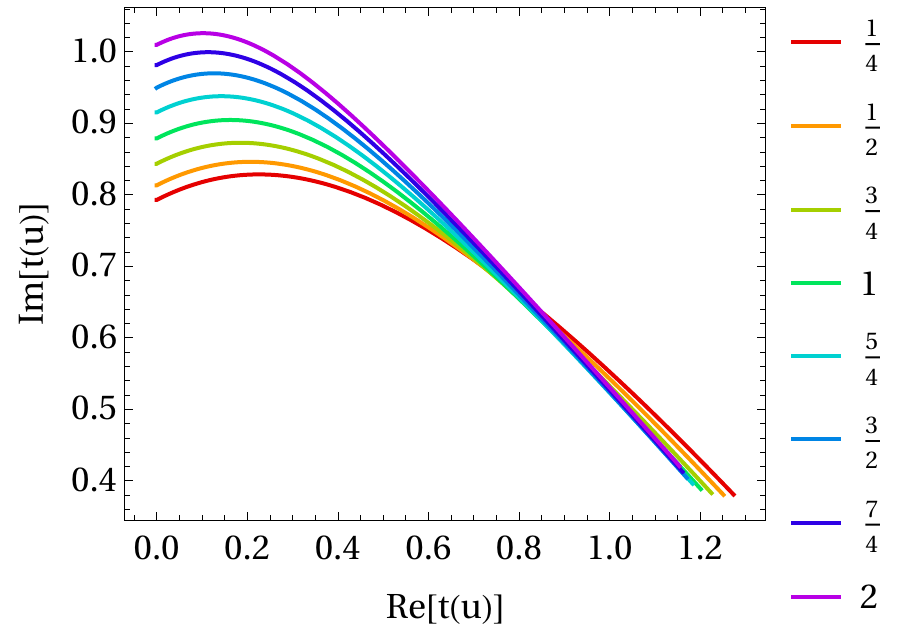}
\vspace{-.4cm}
\caption{Instanton plots for $\gamma_\omega = 1$ and different values of $\gamma_k$. The parametrization is given by $u = e^{-i \theta} r$ with $\theta = 0.2 \pi$.}
\label{fig:instk}
\end{figure}

The parametrization is arbitrary, and all physical quantities are of course independent of this choice. It is possible to choose a parametrization such that $t(r)$ runs parallel to the imaginary axis from the turning point, then turns and becomes purely real. At the momentum saddle point, \eqref{nonTrivialSaddleBoundary} implies that $z$ is also real. We have therefore complex tunneling trajectories at $r \sim 0$ and real particles asymptotically, since both components are real. Such an einbein can be found by setting $\theta = \frac{\pi}{2}$ and tuning the other two parameters until we have the desired result. The resulting instantons are shown in Fig.~\ref{fig:instreal}. Note that the instantons still satisfy
\be
\left(\frac{dq}{du}\right)^2 = 1
\ee
because we have merely changed the contour into $u(r)$. Of course, if we write it in terms of the real parameter $r$ then
\be
\left(\frac{dq}{dr}\right)^2 = f^2(r) \; .
\ee

However, while tuning the einbein such that the instanton goes along the real axis asymptotically may seem natural, from a practical point of view this is actually a great deal of unnecessary work, since we would need to re-tune the einbein whenever we change the parameters of the field or the particle momenta.   
So, for computational convenience we used a simple tilted parametrization as $u(r) = e^{-i \theta} r$ instead, with one fixed value of $\theta$. With the einbein \eqref{eq:Einbein}, we expect the instantons to deviate from the ones with the tilted parametrization only when $\psi$ drops to zero, and this is indeed what happens. 
Instantons for a tilted parametrization are shown in Fig.~\ref{fig:instk}.

\begin{figure}[!ht]
  \centering
\includegraphics[width=\linewidth]{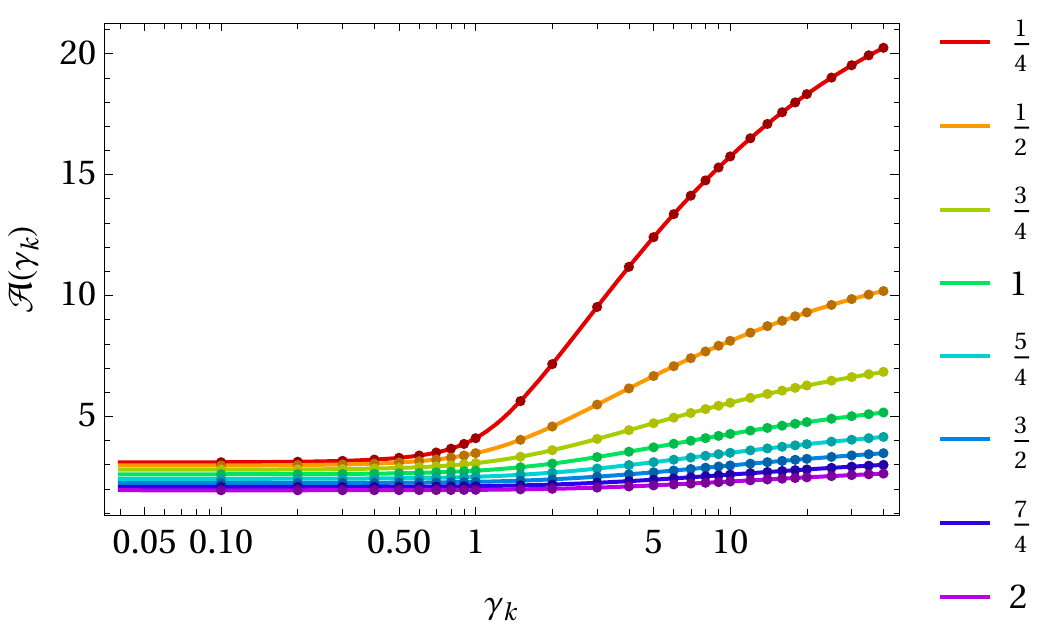}  
\includegraphics[width=\linewidth]{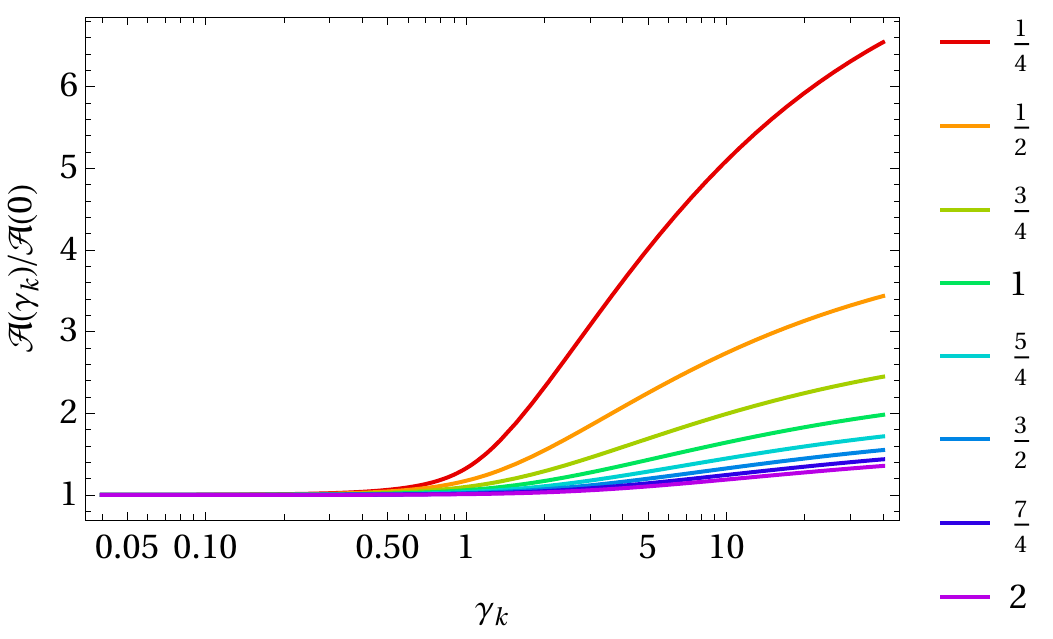}
\caption{Exponential part of the probability $\mathcal A$ from~\eqref{generalFinExp} (without the overall factor of $1/E$) as a function of $\gamma_k$ for various values of $\gamma_\omega$.
The dots are obtained with the discrete-instanton code from~\cite{Schneider:2018huk,Torgrimsson:2017cyb}. The number of points used for the discrete instantons varies depending on the $\gamma_\omega$ value, from $N = 500$ for $\gamma_\omega = 2$ to $N = 2000$ for $\gamma_\omega = \frac{1}{4}$.}
\label{fig:exponent}
\end{figure}

To find the instantons it is convenient to use ``initial'' conditions in the middle of the instanton, at $u=0$, and then vary the position and velocity, $q_\mu(0)$ and $q'_\mu(0)$, until we find a solution with the correct asymptotic momenta at $u\to\pm\infty$. In the symmetric case we can choose $z(0)=0$ and $t'(0)=0$, which implies $z'(0)^2=-m_\LCperp^2$, and then different choices of $t(0)$ lead to different asymptotic momenta.   
$t(0)$ is in general complex. We can determine it using the condition
\be\label{ReImCondition}
z'(u_1) \overset{!}{=} p \;.
\ee
Note that this equation represents two real conditions; one for $\text{Re}(z'(u_1))$ and another for $\text{Im}(z'(u_1))$. \eqref{ReImCondition} allows us to consider a generic momentum $p$, and then we can vary $p$ to find the saddle point $p_s$. However, for $p=p_s$, $t(0)$ turns out to be purely imaginary, so then we only need one real condition to determine $t(0)$. We have found that one possible condition is 
\be\label{ImCondition}
\text{Im} \, z(u_1) \overset{!}{=} 0 \; .
\ee
Note that the momentum $p_s$ does not enter~\eqref{ImCondition}, so we can actually find the dominant instanton without first determining $p_s$. Once we have found this instanton, we can obtain $p_s$ by simply evaluating $p_s=z'(u_1)$. This is significantly faster than using the condition~\eqref{ReImCondition}.

Another thing that can be crucial for finding instantons is to use a numerical continuation~\cite{Schneider:2018huk}, where we start with some parameter values that lead to a simple instanton and then we gradually change the parameters, which leads to a gradual change in the instanton. This idea was used in~\cite{Schneider:2018huk} to find discretized, closed-loop instantons for the imaginary part of the effective action. Here we do not discretize the instanton and our instantons are open lines, but we have still found numerical continuation to be very useful. In particular, if we know the instanton for some value of $\gamma_k$, then we use the value of $t(0)$ as a starting point for the numerical root-finding of $t(0)$ for $\gamma_k + \Delta \gamma_k$. If each step $\Delta \gamma_k$ is sufficiently small, then the root finding converges fast. For a purely time-dependent Sauter pulse we have 
\be\label{tTurn0}
t(0) = \frac{i}{\gamma_\omega} \,  \text{arctan}(\gamma_\omega)
\ee
therefore we use this at the initial point $\gamma_k = 0$. Without this numerical continuation it is difficult to find the instantons at larger $\gamma_k$.

Having obtained the instantons, we can now immediately obtain the results for the exponential part of the probability using~\eqref{generalFinExp}. The results are shown in Fig.~\ref{fig:exponent}. Also shown are the results obtained using the discrete-instanton code from~\cite{Schneider:2018huk,Torgrimsson:2017cyb}. We have perfect agreement.
For $\gamma_\omega\sim1/4$ we see a significant increase in the exponent as $\gamma_k$ increases. In contrast, for $\gamma_\omega\sim2$ the exponent is quite flat, i.e. it is quite insensitive to the spatial width of the field.

\section{Perpendicular momentum integral}\label{sec:PerpIntegral}

To obtain the total probability we can perform the momentum integrals in~\eqref{finalFormula} with the saddle-point method. When doing so we actually obtain the relevant information for the shape of the spectrum near the dominant peaks too, as we will now explain. 

We start with the perpendicular momentum integrals.
Since a nonzero $p_\LCperp$ basically means making the fermions heavier, we have a saddle point at $p_\LCperp=0$. Around this point the spectrum is Gaussian,
\be
\mathbb P(p_\LCperp)\propto\exp\left\{-\frac{p_\LCperp^2}{d_\LCperp^2}\right\} \;.
\ee
To find the width we go back and express the exponent as in~\eqref{LSZ3pair}, but with the integration variables replaced by their saddle-point values. Since this is already a function of $p_\LCperp^2$ rather than $p_\LCperp$, we just have to make a linear expansion. Since the partial derivatives with respect to the previous integration variables vanish at the saddle point, we simply find
%\be
%d_\LCperp^{-2}=-\lim_{T\to\infty}\text{Re }i\left(\frac{t(u_1)}{p_0}+\frac{t(u_0)}{p'_0}-u_1 + u_0 \right) \;,
%\ee
\be\label{dperpEq}
d_\LCperp^{-2}=-\lim_{-u_0,u_1\to\infty}\text{Re }i\left(\frac{t(u_1)}{p_0}+\frac{t(u_0)}{p'_0}-u_1 + u_0 \right) \;.
\ee
We can see that this limit is finite by by noting that the derivatives with respect to $u_0$ and $u_1$ vanish outside the field. We can therefore choose any $u_0$ and $u_1$ that are large enough so that the instanton at these points is outside the field. Interestingly, note that, by using~\eqref{dperpEq}, we obtain the transverse width $d_\LCperp$ from instantons with $p_\LCperp=0$, i.e. we do not need to find any instantons with $p_\LCperp\ne0$ for this.
From~\eqref{dperpEq}, \eqref{rescaleqwithE} and~\eqref{rescaleuwithE} we see that $d_\LCperp\propto\sqrt{E}$.

\begin{figure}[!ht]
\centering
\includegraphics[width=\linewidth]{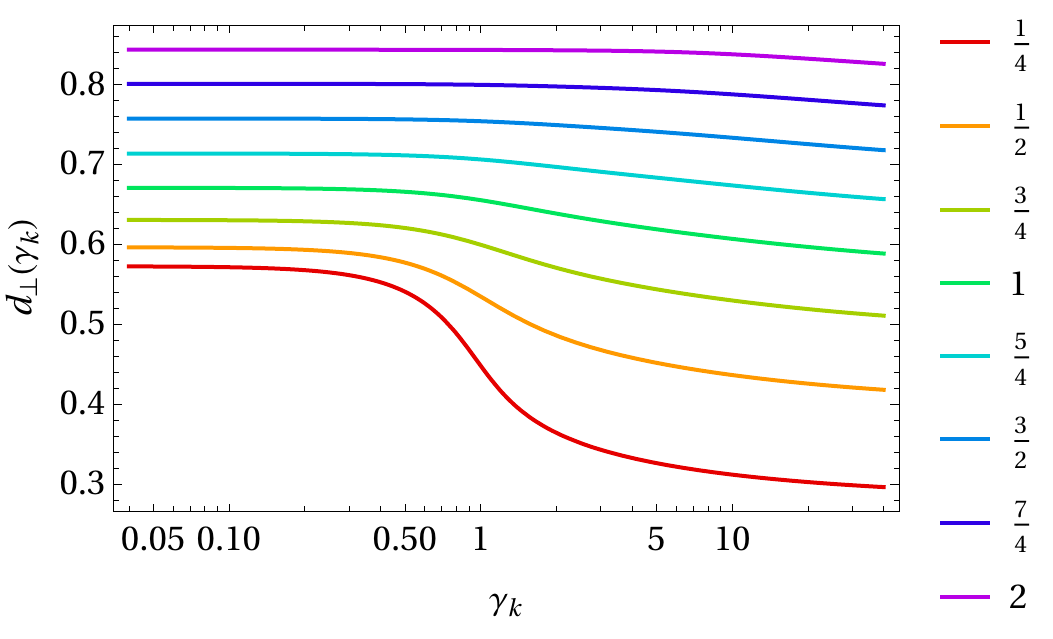}
\includegraphics[width=\linewidth]{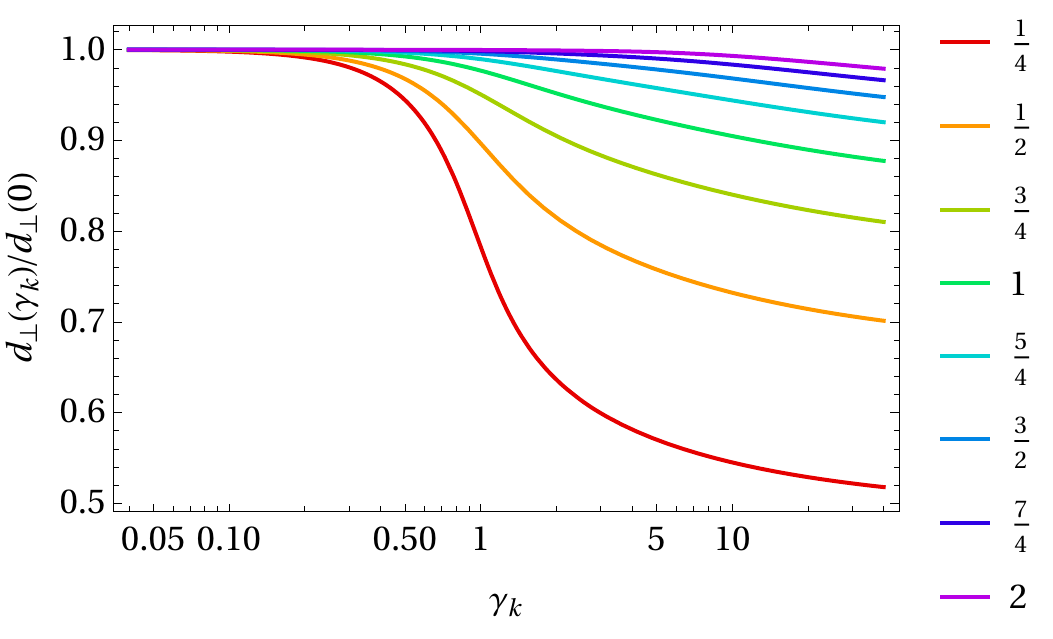}
\caption{$d_\perp(\gamma_k)$ (without the overal factor of $\sqrt{E}$) for different values of $\gamma_\omega$, calculated using~\eqref{dperpEq}.}
\label{fig:dPerp}
\end{figure}

The results are shown in Fig.~\ref{fig:dPerp}. We see that both the exponent in Fig.~\ref{fig:exponent} and $d_\perp$ in Fig.~\ref{fig:dPerp} are, for $\gamma_\omega\gtrsim1$, quite insensitive to $\gamma_k$.

\section{Longitudinal momentum integral for symmetric fields}\label{sec:LongIntegral}

For a space-independent field we would have a delta function $\delta(p_3+p'_3)$, which we do not have for the space-time dependent fields we now consider. However, for symmetric fields we still have a saddle point at $p_3+p'_3=0$. We therefore change variables from
\be
p_3=-P+\frac{\Delta p}{2}
\qquad
p'_3=P+\frac{\Delta p}{2}
\ee
to $P$ and $\Delta p$. It follows from the symmetry that there is a saddle point at $\Delta p=0$, regardless of the value of $P$. 

\subsection{The component with trivial saddle point}

To obtain the Gaussian width $d_\Delta$ in 
\be
\mathbb P (\Delta p)\propto\exp\left\{-\frac{\Delta p^2}{d_\Delta^2}\right\}
\ee 
we start with the exponent expressed as in~\eqref{LSZ3pair}, where all the integration variables have been replaced by their saddle-point values. Since the partial derivatives of the exponent with respect to the integration variables vanish at the saddle point, we have
\be\label{dAdp}
\frac{\partial\mathcal{A}}{\partial p_3}=-2\text{Re }i\left[z(u_1)+\frac{p_3}{p_0}t(u_1)\right]
\ee
and
\be\label{dAdpp}
\frac{\partial\mathcal{A}}{\partial p'_3}=-2\text{Re }i\left[z(u_0)+\frac{p'_3}{p'_0}t(u_0)\right] \;.
\ee
As the width is given by
\be
d_\Delta^{-2}=\frac{1}{2}\mathcal{A}''(\Delta p=0) \;,
\ee
we obtain it by expanding~\eqref{dAdp} and~\eqref{dAdpp} to linear order in $\Delta p$.
So, we only need the first-order variation of the instanton,
\be
q_\mu(u)\to q_\mu(u)+\Delta p\, \delta q_\mu(u) \;.
\ee
It follows from~\eqref{z0saddleEq} and~\eqref{z1saddleEq} that
\be\label{tAsymptoticP}
-t'(u_0)=t'(u_1)=p_0
\qquad
z'(u_0)=z'(u_1)=P
\ee
\be\label{trivialSaddleBoundary}
\delta t'(u_0)=\delta t'(u_1)=-\frac{P}{2p_0}
\qquad
\delta z'(u_0)=-\delta z'(u_1)=\frac{1}{2} \;,
\ee
where $p_0=\sqrt{m_\LCperp^2+P^2}$.
The expansion of the Lorentz force equation gives
\be\label{deltaqeqs}
\begin{split}
\delta t''&=\nabla E\cdot\{\delta t,\delta z\}z'+E\delta z'\\
\delta z''&=\nabla E\cdot\{\delta t,\delta z\}t'+E\delta t' \;.
\end{split}
\ee
From~\eqref{trivialSaddleBoundary} we find
\be\label{quad0}
\begin{split}
d_\Delta^{-2}&=\frac{1}{2}\mathcal{A}''(\Delta p=0)\\
&=\text{Re }i\left\{-\frac{m_\LCperp^2}{2p_0^3}t+\frac{P}{p_0}\delta t-\delta z\right\}\bigg|_{u\to\infty} \;.
\end{split}
\ee 

We can express this in terms of $\eta=t'\delta z-z'\delta t$ as
\be\label{dDeltaFrometa}
d_\Delta^{-2}=\text{Re }\frac{im_\LCperp^2}{2p_0^2}\left(\eta_a-\frac{t}{p_0}\right)\bigg|_{u\to\infty} \;,
\ee
where
\be
\eta=-\frac{m_\LCperp^2}{2p_0}\eta_a 
\ee
is a solution to~\eqref{etaEq} with $\nu=0$ and
\be\label{etaDeltaBoundary}
\eta_a'(u_0)=\eta_a'(u_1)=1 \;.
\ee
$\eta_a$ is therefore antisymmetric in $u$.
When rescaling as in~\eqref{rescaleqwithE} and~\eqref{rescaleuwithE}, we see from~\eqref{etaDeltaBoundary} that $\eta_a\propto1/E$, and so from~\eqref{dDeltaFrometa} it follows that $d_\Delta\propto\sqrt{E}$.

\begin{figure}[!ht]
\centering
\includegraphics[width=\linewidth]{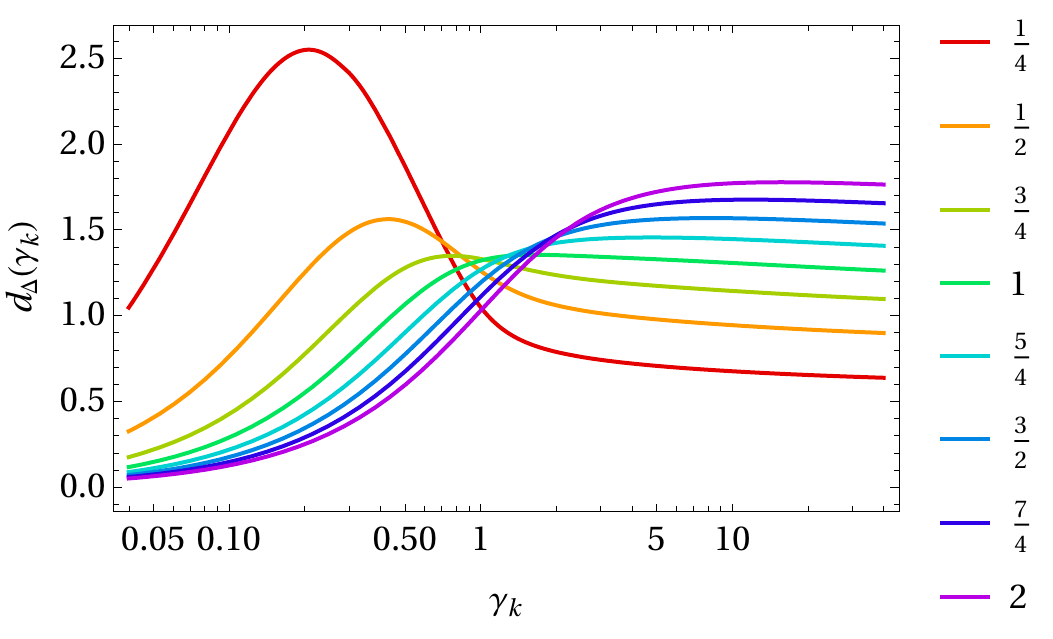}
\includegraphics[width=\linewidth]{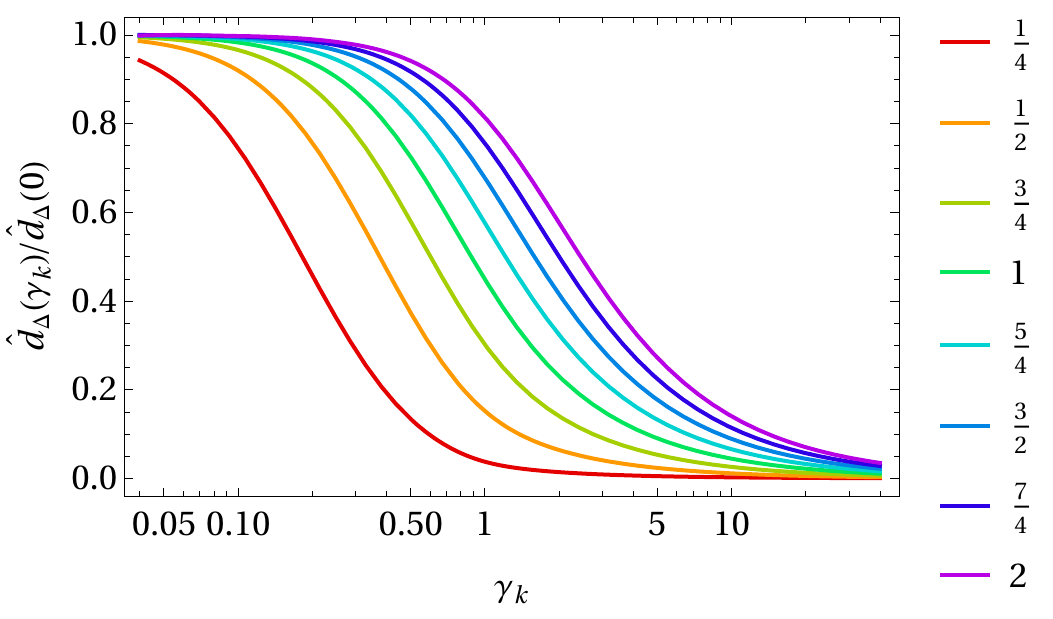}
\caption{$d_\Delta(\gamma_k)$ (without the overall factor of $\sqrt{E}$) for different values of $\gamma_\omega$, calculated using~\eqref{dDeltaFrometa} or~\eqref{quad0}. Here $d_\Delta = \gamma_k \hat d_\Delta$.}
\label{fig:dDelta}
\end{figure}

The results are shown in Fig.~\ref{fig:dDelta}. In contrast to Fig.~\ref{fig:exponent} and Fig.~\ref{fig:dPerp}, Fig.~\ref{fig:dDelta} shows that $d_\Delta$ is much more sensitive to the value of $\gamma_k$. 

\subsection{The component with nontrivial saddle point}

Now that the $\Delta p$ integral has been performed we have
\be
p_3= -P \qquad p'_3=P \;.
\ee
To obtain the width in $P$ we can use the same method as for $\Delta p$. We write $P\to P+\delta P$, where afterwards $P$ now denotes the saddle-point value, which is nontrivial, and $\delta P$ is the new integration variable. We again expand the instanton to linear order
\be
q_\mu(u)\to q_\mu(u)+\delta P\, \delta q_\mu(u) \;.
\ee
The equations of motion for $\delta q_\mu$ are the same, i.e.~\eqref{deltaqeqs}, and we can still use~\eqref{dAdp} and~\eqref{dAdpp}. The difference is the boundary/initial conditions,
\be\label{nonTrivialSaddleBoundary}
\delta z'(u_0)=\delta z'(u_1)=1
\qquad
-\delta t'(u_0)=\delta t'(u_1)=\frac{P}{p_0} \; .
\ee
So this time $\delta t$ is symmetric, while $\delta z$ is antisymmetric.

The saddle point for $P$ is determined by
\be\label{PsaddleEq}
\mathcal{A}'(P)=-4\text{Re }i\left[\frac{P}{p_0}t(u_1)-z(u_1)\right]\overset{!}{=}0 \;,
\ee
where we have used the symmetry to set $t(u_0)=t(u_1)$ and $z(u_0)=-z(u_1)$. This is independent of $u_1$ (as long as it is chosen sufficiently large), which follows from the asymptotic values of $t'$ and $z'$.

\begin{figure}[!ht]
\centering
\includegraphics[width=\linewidth]{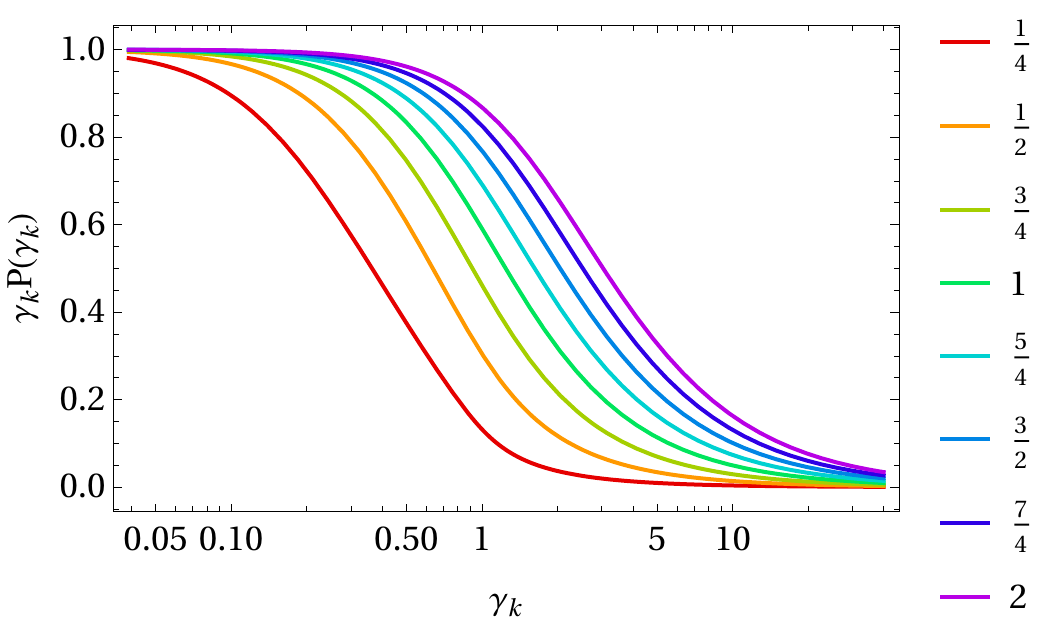}
\vspace{-.8cm}
\caption{Normalized saddle-point value of $P$ for fixed $\gamma_\omega$ as function of $\gamma_k$. We always have $P(0) = 1/\gamma_\omega$.}
\label{fig:P(k)}
\end{figure}

$P$ for the spacetime Sauter pulse is plotted in Fig.~\ref{fig:P(k)}. We see that $P$ decreases as $\gamma_k$ grows, since the pairs are more likely to be produced with smaller momenta when the size of the field gets smaller in the $z$ direction.

Using the asymptotic boundary conditions in~\eqref{nonTrivialSaddleBoundary} we find
\be\label{dDeltaEq}
d_P^{-2}=\frac{1}{2}\mathcal{A}''(\delta P=0)=
-2\text{Re }i\left\{\frac{m_\LCperp^2}{p_0^3}t+\frac{P}{p_0}\delta t-\delta z\right\}\bigg|_{u\to\infty} \;.
\ee
This too can be written in terms of $\eta$,
\be\label{dPfrometa}
d_P^{-2}=\text{Re }\frac{2im_\LCperp^2}{p_0^2}\left(\eta_s-\frac{t}{p_0}\right)\bigg|_{u\to\infty} \;,
\ee
where 
\be
\eta=\frac{m_\LCperp^2}{p_0}\eta_s
\ee
is a solution to~\eqref{etaEq} with $\nu=0$ and
\be\label{etaPBoundary}
-\eta_s'(u_0)=\eta_s'(u_1)=1 \;.
\ee
$\eta_s$ is therefore symmetric in $u$.
When rescaling as in~\eqref{rescaleqwithE} and~\eqref{rescaleuwithE}, we see from~\eqref{etaPBoundary} that $\eta_s\propto1/E$, and so from~\eqref{dPfrometa} it follows that $d_P\propto\sqrt{E}$.

\begin{figure}[!ht]
\centering
\includegraphics[width=\linewidth]{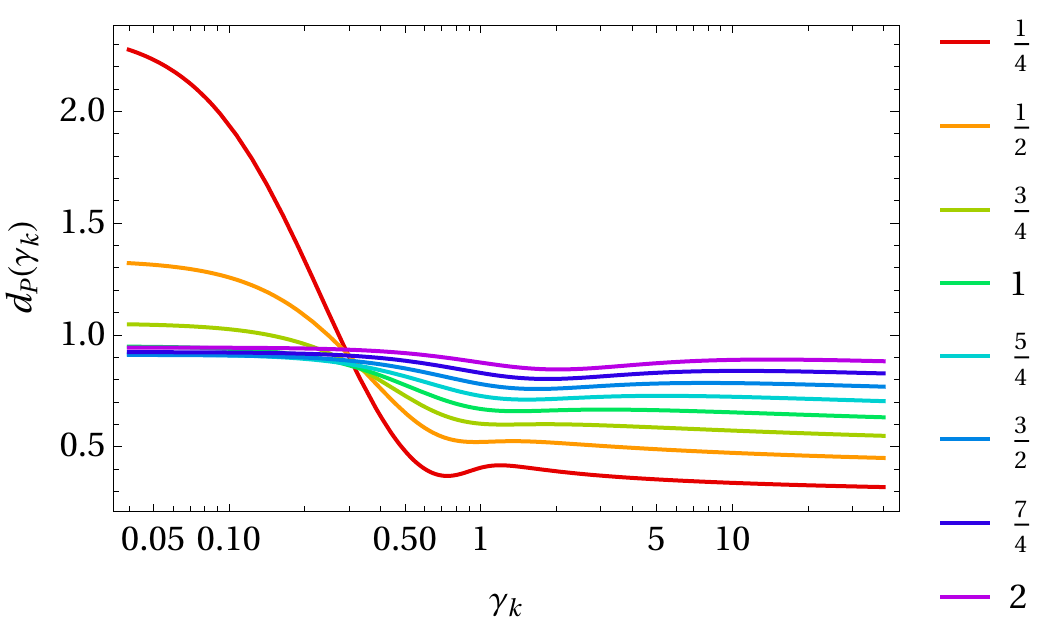}
\includegraphics[width=\linewidth]{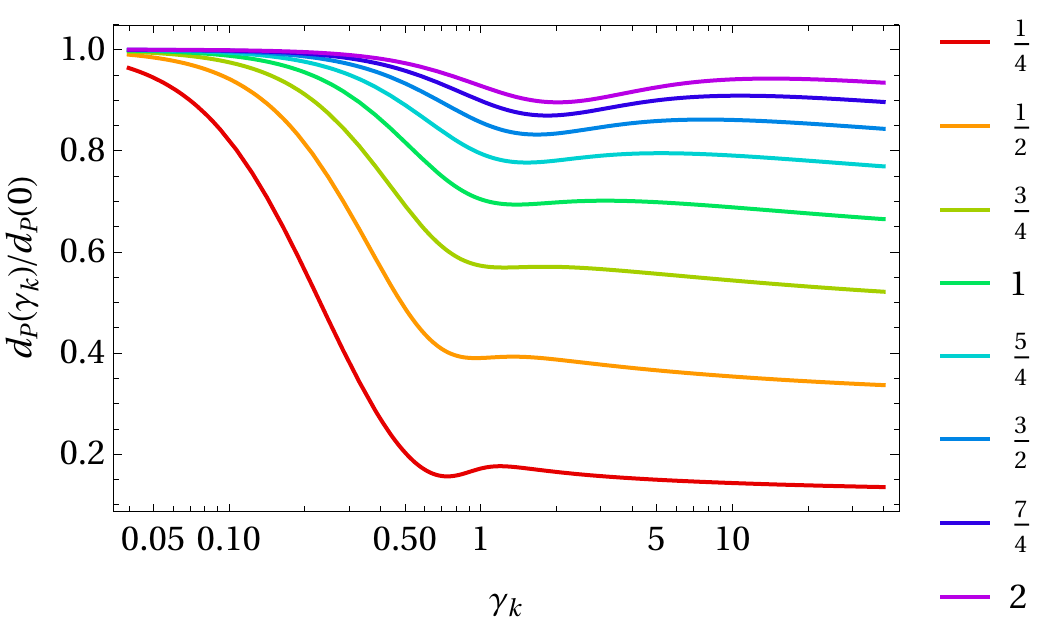}
\caption{$d_P(\gamma_k)$ (without the overall factor of $\sqrt{E}$) for different values of $\gamma_\omega$, calculated using~\eqref{dPfrometa} or~\eqref{dDeltaEq}.}
\label{fig:dP}
\end{figure}

The results for $d_P$ are shown in Fig.~\ref{fig:dP}.

\subsection{$\eta$ solutions}

Note that we have now expressed the contribution from the functional determinant (see~\eqref{hFrometa}), $d_\Delta$~\eqref{dDeltaFrometa} and $d_P$~\eqref{dPfrometa} in terms of solutions of~\eqref{etaEq} with $\nu=0$. The difference between these three contributions is the initial/boundary conditions in~\eqref{etaInitialForh},\eqref{etaDeltaBoundary} and~\eqref{etaPBoundary}. Since there are only two linearly independent solutions to~\eqref{etaEq}, we can write e.g. the $\eta$ solution with~\eqref{etaInitialForh} as a superposition of the symmetric and antisymmetric solutions
\be
\eta(u)=c_a\eta_a(u)+c_s\eta_s(u) \;.
\ee
From~\eqref{etaInitialForh},\eqref{etaDeltaBoundary} and~\eqref{etaPBoundary} we find 
\be
c_a=c_s=\frac{1}{\eta_s(u_1)-\eta_a(u_1)} \;.
\ee
This means that we can write the contribution from the functional determinant in terms of the same combinations that appear in~\eqref{dDeltaFrometa} and~\eqref{dPfrometa} for $d_\Delta$ and $d_P$ as
\be\label{hWidths}
h(u_1)=2\left[\left(\eta_s-\frac{t}{p_0}\right)-\left(\eta_a-\frac{t}{p_0}\right)\right]^{-1}\bigg|_{u\to\infty} \;.
\ee

In fact, once we have found one solution to~\eqref{etaEq} ($\eta_a$ say) then the other one ($\eta_s$) can be obtained using Abel's identity, as explained in Appendix~\ref{AbelSection}.

\section{The $k\to0$ limit}\label{sec:kToZero}

In this section we will consider the limit where the field depends very slowly on $z$, i.e. $k\to0$.
In some contributions to the probability we can simply set $k=0$. However, for a space-independent field we would have a delta function in $\delta(p_3+p'_3)$, so we expect that the Gaussian width for $\Delta p$ should become increasingly narrow, i.e. $d_\Delta\to0$. 
We also find that $h\to0$ (the contribution from the path integral, see~\eqref{finalFormula}). Thus for $d_\Delta$ and $h$ we have to derive nontrivial $k\ll1$ approximations in order to obtain the probability to leading order in $k\ll1$.

We have found that we can obtain such approximations by making a power-series expansion in $k^2$.  
We obtain $h$ from~\eqref{etaEq} with $\nu=0$ going up to next-to-leading order of the expansion of $\eta$ in $k^2$. The limit of $d_\Delta$ follows immediately after we have found $h$.
To obtain this we first need to find the first two terms in the $k^2$ expansion of the instanton. 

\subsection{Instanton}

It turns out that the instanton can be expanded as
\be
q^\mu\approx q_{(0)}^\mu+k^2 q_{(1)}^\mu \;.
\ee
To zeroth order we have
\be\label{t0z0A0}
t_{(0)}'=\pm\sqrt{m_\LCperp^2+A_{(0)}^2}
\qquad
z_{(0)}'=A_{(0)} \;,
\ee
where $A_{(0)}(t_{(0)})=A(k\to0)$. For a Sauter pulse we have $A_{(0)}(t_{(0)})=(E/\omega)\tanh(\omega t_{(0)})$. 
This gives an implicit equation for $t_{(0)}'$ in terms of $t_{(0)}$, but at the turning point $t_{(0)}'(0)=0$ we find an explicit expression for $t_{(0)}(0)$. Setting $p_\LCperp=0$ we find
\be
t_{(0)}(0)=\frac{i}{\omega}\arctan(\gamma_\omega) \;,
\ee
which agrees with~\eqref{tTurn0} after the rescaling in~\eqref{rescaleqwithE}. 
The asymptotic momentum of this is fixed, i.e.
\be
z_{(0)}'(u_0)=z_{(0)}'(u_1)=\frac{1}{\gamma_\omega} \;,
\ee
so for this to be consistent with~\eqref{tAsymptoticP} we see that the saddle-point value of the longitudinal momentum has to scale as
\be
P\approx\frac{1}{\gamma_\omega}-c k^2
\ee 
for $k\ll1$, and where $c$ is a constant. The leading order, $1/\gamma_\omega$, agrees with the numerical results in Fig.~\ref{fig:P(k)}. Thus, the boundary conditions for the next-to-leading order is
\be\label{z1Asymptotic}
z_{(1)}'(u_0)=z_{(1)}'(u_1)=-c
\ee
and the equations of motion
\be
\begin{split}
t_{(1)}''&=E_{(0)}z_{(1)}'+(E_{(0)}'t_{(1)}-\zeta E_{(0)}z_{(0)}^2)z_{(0)}' \\
z_{(1)}''&=E_{(0)}t_{(1)}'+(E_{(0)}'t_{(1)}-\zeta E_{(0)}z_{(0)}^2)t_{(0)}' \;,
\end{split}
\ee
where $E_{(0)}(t_{(0)})=E(k\to0)$, $E_{(0)}'=\ud E_{(0)}/\ud t$ and $\zeta$ is defined by
\be
E(t,z)=E_{(0)}(t)(1-\zeta(kz)^2+\mathcal{O}([kz]^4)) \;.
\ee
For a Sauter pulse we have $E_{(0)}(t)=E\text{sech}^2(\omega t_{(0)})$ and $\zeta=1$. 
Similarly to what we did in Sec.~\ref{sec:Instantons}, we can actually solve for $q_{(1)}$ without knowing the constant $c$ in~\eqref{z1Asymptotic}. We do this by setting the conditions
\be
z_{(1)}(0) = z^\prime_{(1)}(0) = t^\prime_{(1)}(0) = 0
\ee
and vary the purely imaginary value $t(0)$ until
\be
\text{Im} \, z_{(1)}(u_1) \overset{!}{=} 0 \; .
\ee
After we have found the solution, we can find $c$ by simply evaluating $c=-z_{(1)}'(u_1)$.

\subsection{Functional determinant and $d_\Delta$}

To obtain the $k \to 0$ limit of the Gelfand-Yaglom determinant \eqref{hFrometa} we perform a Taylor expansion 
\be
\eta(u)\to\eta_{(0)}+k^2\eta_{(1)} \;
\ee
and observe that the equations for these two terms are given by
\be\label{etaEqSmallk}
\begin{split}
\eta_{(0)}'' &= \left(E^2_{(0)}(t_{(0)})+E'_{(0)}(t_{(0)})z'_{(0)}\right) \eta_{(0)} \\
\eta_{(1)}'' &= 
\left(E^2_{(0)}(t_{(0)}) + z'_{(0)} \, E'_{(0)}(t_{(0)}) \right) \eta_{(1)} \\
&+\biggr(2E_{(0)}(t_{(0)})\left[E'_{(0)}(t_{(0)})t_{(1)} - \zeta E_{(0)}(t_{(0)}) z_{(0)}^2 \right] \\ 
&-2\zeta z_{(0)} E_{(0)}(t_{(0)}) t'_{(0)}+ E''_{(0)}(t_{(0)}) t_{(1)} z'_{(0)} \\
&+ E'_{(0)}(t_{(0)}) z'_{(1)} - \zeta z_{(0)}^2 E'_{(0)}(t_{(0)}) z'_{(0)} \biggr)\eta_{(0)} \; .
\end{split}
\ee
With the initial conditions~\eqref{etaInitialForh} we immediately find
\be\label{eta0forh}
\eta_{(0)}(u)=-\frac{t_{(0)}^\prime}{p_0}
\ee
and for $\eta_{(1)}$ we solve \eqref{etaEqSmallk} with initial conditions
\be
\eta_{(1)}(\tilde u_0) = \eta_{(1)}^\prime(\tilde u_0) = 0 \; .
\ee
Since $\eta_{(0)}'(\tilde u_1)=0$, the limit $k \to 0$ of $h(\tilde u_1)$ is simply
\be
h(\tilde{u}_1)\approx k^2\eta_{(1)}^\prime(\tilde{u}_1) \;.
\ee

As for $d_\Delta$, the simplest way to obtain it is from~\eqref{dDeltaFrometa} using~\eqref{hWidths}. Since $d_P$ is finite when $k \to 0$ we simply have
\be
d_\Delta^{-2} = \text{Im} \, \frac{1}{\left(1+\frac{1}{\gamma_\omega^2}\right) h(\tilde u_1)} \; .
\ee
Thus, we see that $d_\Delta\propto k$ indeed goes to zero as $k\to0$.

\subsection{Total prefactor}

In summary, in the $k\to0$ limit we have $d_\Delta= k\hat{d}_\Delta$ and $h=k^2\hat{h}$, so the total prefactor scales as
\be
\mathbb P(p)\propto\frac{1}{2\pi k^2\hat{h}}\exp\left\{-\frac{\Delta p^2}{k^2\hat{d}_\Delta^2}\right\}\to\frac{\hat{d}_\Delta}{2\sqrt{\pi}k\hat{h}}\delta(\Delta p) \;.
\ee
This factor would be $V_z\delta(\Delta p)$ had we started with $k=0$, where $V_z$ is a volume factor. Here we find a regularized volume factor proportional to $1/k$.

%\section{Numerical results}

\begin{figure}[!ht]
\centering
\includegraphics[width=\linewidth]{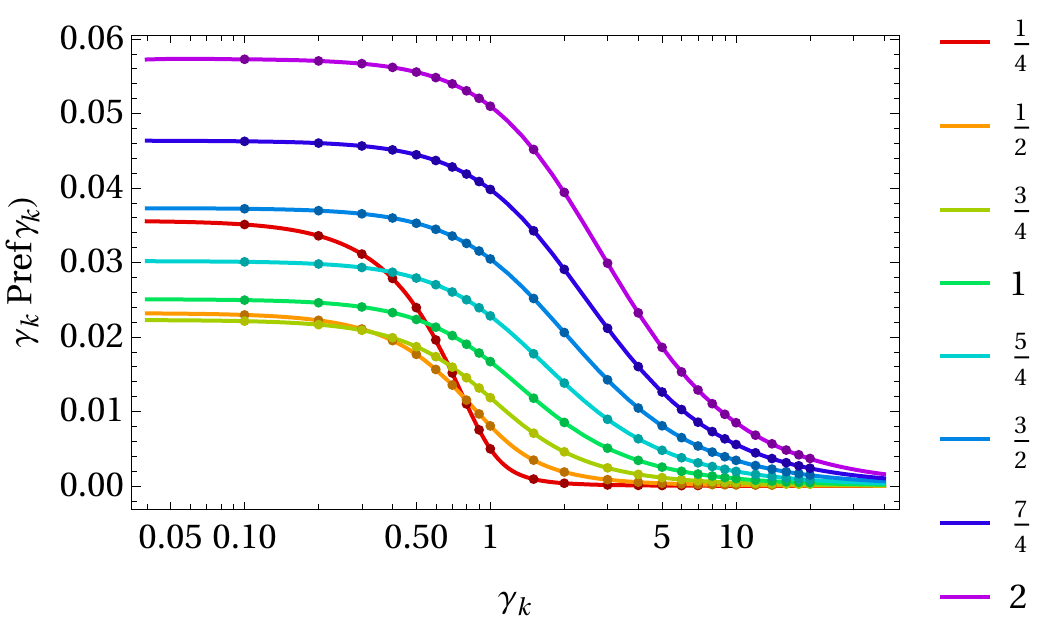}
\includegraphics[width=\linewidth]{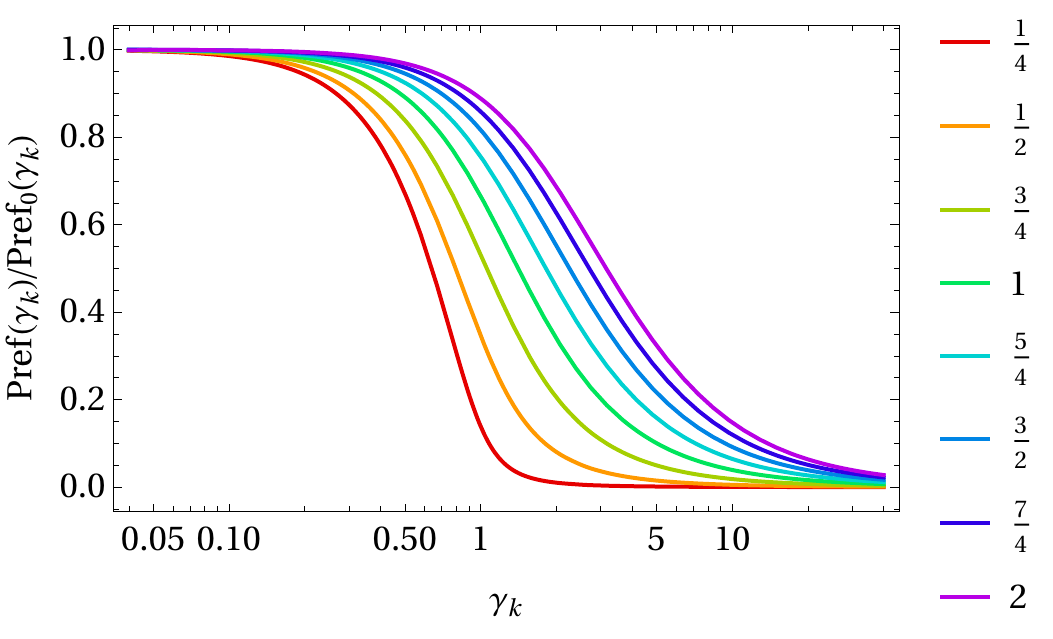}
\caption{Prefactor $\mathrm{Pref}(\gamma_k)$ (without the overall factor of $E$) for different values of $\gamma_\omega$. We have divided by either $1/\gamma_k$ or $\mathrm{Pref}_0(\gamma_k)$ to remove the $1/k$ scaling. The dots are obtained with the code in~\cite{Schneider:2018huk,Torgrimsson:2017cyb}. The number of points used for the discrete instantons varies depending on the $\gamma_\omega$ value from $N = 2000$ for $\gamma_\omega = 2$ to $N = 4000$ for $\gamma_\omega = \frac{1}{4}$.}
\label{fig:prefactor}
\end{figure}

We plot the normalized probability without the factor of $1/k$ by dividing by the leading order contribution as $\gamma_k \to 0$
\be\label{pref0}
\mathrm{Pref}_0(k) := \lim_{k \to 0} \mathrm{Pref}(k)\propto\frac{1}{k} \; .
\ee
The results for the Sauter pulse is shown in Fig.~\ref{fig:prefactor}. The plot also demonstrates perfect agreement with the results obtained with the instanton code from~\cite{Schneider:2018huk,Torgrimsson:2017cyb}, which deals with closed instantons for the imaginary part of the effective action. This prefactor is the product of several different contributions in our new approach, it is in particular a product of the widths of the spectrum, so this agreement is not just a check for the integrated probability but also for the spectrum, which one cannot obtain with the closed-instanton approach.

\section{Conclusions}
In this paper we developed a method that makes use of worldline instantons with open lines to obtain the pair-production spectrum in the presence of a background field which depends not only on time, as in previous works, but also on one space coordinate. 
To do so we made use of the LSZ reduction formula with free asymptotic states and internal propagator expressed in its worldline representation with a particle path integrals.
From the spectrum, we showed how the maximum changes with the field shape. In particular, as one might expect, when the spatial extension of the field gets smaller, it is more likely to produce particles with smaller momenta. Since the field depends on one spatial dimension, the momentum is not conserved along that direction. Nonetheless, the spectrum is symmetric under electron/positron exchange. 

From the integrated spectrum we could also obtain the total probability, finding perfect agreement with results obtained using the discrete-instanton code from~\cite{Schneider:2018huk,Torgrimsson:2017cyb}. 

This method should also work for fields which depend on more spatial directions and have magnetic components. For example, an interesting and more realistic field is the e-dipole pulse \cite{Gonoskov:2012}, which is a solution to Maxwell's field equations localized in all 4 spacetime coordinates. This field was considered in~\cite{Schneider:2018huk} using the closed-instanton approach to obtain the total probability. Now with our new, open-instanton approach we could also study the corresponding momentum spectrum or the spin dependence. 

We showed in~\cite{DegliEsposti:2021its} how to use open instantons for nonlinear Breit-Wheeler pair production and nonlinear Compton scattering in a time dependent field. We expect that these methods can also be used for such processes in spacetime dependent fields. 

The instanton and the usual WKB methods should give equivalent results for the semiclassical approximation, and in the 1D cases where it has been possible to use both this has been confirmed. However, while WKB methods are more well known, and usually easier to use for 1D problems, the instanton approach seems more promising when going beyond 1D field. 

\acknowledgements
We are very grateful to Christian Schneider for providing us with the Mathematica code that he wrote for~\cite{Schneider:2018huk,Torgrimsson:2017cyb}. We have used this code in order to compare the results obtained with the worldline instanton approach to the imaginary part of the effective action, i.e. the total/integrated pair production probability, with our new worldline instanton approach for the momentum spectrum. G.~T. is supported by the Swedish Research Council, contract 2020-04327.

\appendix

\section{Abel's identity}\label{AbelSection}

In this appendix we will explain how to obtain e.g. $\eta_s$ from $\eta_a$ using Abel's identity. In this case it says that, since there is no $\eta'$ term in~\eqref{etaEq}, the Wronskian is constant,
\be
W=\eta_s\eta_a'-\eta_s'\eta_a=\text{const.} \;,
\ee
which can be solved for $\eta_s$ in terms of $\eta_a$. For $u<0$ we can write the solution as
\be\label{etasPole}
\eta_s(u)=\eta_a(u)\left(\frac{W}{\eta_a(u_0)}-1-W\int_{u_0}^u\frac{\ud v}{\eta_a^2(v)}\right)
\ee
The limit of~\eqref{etasPole} as $u\to0$ from $u<0$ remains finite despite the pole in the integrand due to $\eta_a\to0$. However, we cannot directly evaluate~\eqref{etasPole} for $u>0$, since then the integral would go over the pole (and the overall factor of $\eta_a(u)\ne0$). In principle this is not a big problem since $\eta_s$ for $u>0$ follows from the symmetry $\eta_s(u)=\eta_s(-u)$. But it is nevertheless useful to rewrite~\eqref{etasPole} by making a partial integration as
\be
\eta_s(u)=\frac{W}{\eta_a'(u)}+\eta_a(u)\left(-1+W\int_{u_0}^u\ud v\frac{\eta_a''}{\eta_a\eta_a^{\prime2}}\right) \;,
\ee 
where the integrand no longer has a pole. We can find $W$ by demanding that $\eta_s'(0)=0$ ($\eta_s'$ is antisymmetric). We find
\be
\frac{1}{W}=\int_{u_0}^0\ud v\frac{\eta_a''}{\eta_a\eta_a^{\prime2}} \;.
\ee
With this we can now write $\eta_s$ in a manifestly symmetric form,
\be
\eta_s(u)=W\left(\frac{1}{\eta_a'(u)}+\eta_a(u)\int_0^u\ud v\frac{\eta_a''}{\eta_a\eta_a^{\prime2}}\right) \;.
\ee
However, at the end we actually only need $\eta_s(u_1)$ (we use $u_0=-u_1$), which we obtain most easily by going back to~\eqref{etasPole},
\be
\eta_s(u_1)=\eta_s(u_0)=W+\eta_a(u_1) \;.
\ee
In particular, for the functional determinant we have
\be
h(u_1)=\frac{2}{W} \;.
\ee

Here we have singled out $\eta_a$ as the solution in terms of which the other solutions are expressed. This is motivated by the fact that for $k\to0$ it has a simple form. But for numerical purposes it might be more convenient to instead use a solution that is fixed by the value of $\eta$ and $\eta'$ at one point, e.g. as in~\eqref{etaInitialForh}, rather than at two points as in~\eqref{etaDeltaBoundary}, because with~\eqref{etaInitialForh} we only have to solve~\eqref{etaEq} once, while if we find the solution with~\eqref{etaDeltaBoundary} by varying $\eta'(0)$ we would have to solve~\eqref{etaEq} several times until we found the value of $\eta'(0)$ that gives the solution. Thus, for numerical purposes it can be faster to write $\eta_a$ as a superposition of two solutions as
\be
\eta_a(u)=-\frac{\eta_2(u_1)}{1+\eta_1(u_1)}\eta_1(u)+\eta_2(u) \;,
\ee 
where $\eta_1$ has initial conditions as in~\eqref{etaInitialForh} while
\be
\eta_2(\tilde{u}_0)=0
\qquad
\eta_2'(\tilde{u}_0)=1 \;.
\ee
However, for the cases we consider here, \eqref{etaEq} is solved quickly regardless of which approach we use.


\begin{thebibliography}{0}%
\makeatletter
\providecommand \@ifxundefined [1]{%
 \@ifx{#1\undefined}
}%
\providecommand \@ifnum [1]{%
 \ifnum #1\expandafter \@firstoftwo
 \else \expandafter \@secondoftwo
 \fi
}%
\providecommand \@ifx [1]{%
 \ifx #1\expandafter \@firstoftwo
 \else \expandafter \@secondoftwo
 \fi
}%
\providecommand \natexlab [1]{#1}%
\providecommand \enquote  [1]{``#1''}%
\providecommand \bibnamefont  [1]{#1}%
\providecommand \bibfnamefont [1]{#1}%
\providecommand \citenamefont [1]{#1}%
\providecommand \href@noop [0]{\@secondoftwo}%
\providecommand \href [0]{\begingroup \@sanitize@url \@href}%
\providecommand \@href[1]{\@@startlink{#1}\@@href}%
\providecommand \@@href[1]{\endgroup#1\@@endlink}%
\providecommand \@sanitize@url [0]{\catcode `\\12\catcode `\$12\catcode
  `\&12\catcode `\#12\catcode `\^12\catcode `\_12\catcode `\%12\relax}%
\providecommand \@@startlink[1]{}%
\providecommand \@@endlink[0]{}%
\providecommand \url  [0]{\begingroup\@sanitize@url \@url }%
\providecommand \@url [1]{\endgroup\@href {#1}{\urlprefix }}%
\providecommand \urlprefix  [0]{URL }%
\providecommand \Eprint [0]{\href }%
\providecommand \doibase [0]{http://dx.doi.org/}%
\providecommand \selectlanguage [0]{\@gobble}%
\providecommand \bibinfo  [0]{\@secondoftwo}%
\providecommand \bibfield  [0]{\@secondoftwo}%
\providecommand \translation [1]{[#1]}%
\providecommand \BibitemOpen [0]{}%
\providecommand \bibitemStop [0]{}%
\providecommand \bibitemNoStop [0]{.\EOS\space}%
\providecommand \EOS [0]{\spacefactor3000\relax}%
\providecommand \BibitemShut  [1]{\csname bibitem#1\endcsname}%
\let\auto@bib@innerbib\@empty
%</preamble>
\end{thebibliography}%


\begin{thebibliography}{99}

%\cite{Sauter:1931}
\bibitem{Sauter:1931}
F.~Sauter,
``\"Uber das Verhalten eines Elektrons im homogenen elektrischen Feld nach der relativistischen Theorie Diracs,''
Z. Phys. \textbf{69} (1931) 742

%\cite{Schwinger:1931}
\bibitem{Schwinger:1951}
J.~S.~Schwinger,
``On gauge invariance and vacuum polarization,''
Phys. Rev. \textbf{82} (1951) 664.



%\cite{Ilderton:2014mla}
\bibitem{Ilderton:2014mla}
A.~Ilderton,
``Localisation in worldline pair production and lightfront zero-modes,''
JHEP \textbf{09}, 166 (2014)
%doi:10.1007/JHEP09(2014)166
[arXiv:1406.1513 [hep-th]].
%40 citations counted in INSPIRE as of 16 Dec 2022

%\cite{Breev:2021lpn}
\bibitem{Breev:2021lpn}
A.~I.~Breev, S.~P.~Gavrilov, D.~M.~Gitman and A.~A.~Shishmarev,
``Vacuum instability in time-dependent electric fields: New example of an exactly solvable case,''
Phys. Rev. D \textbf{104}, no.7, 076008 (2021)
%doi:10.1103/PhysRevD.104.076008
[arXiv:2106.06322 [hep-th]].
%4 citations counted in INSPIRE as of 19 Dec 2022

%\cite{Aleksandrov:2017mtq}
\bibitem{Aleksandrov:2017mtq}
I.~A.~Aleksandrov, G.~Plunien and V.~M.~Shabaev,
``Momentum distribution of particles created in space-time-dependent colliding laser pulses,''
Phys. Rev. D \textbf{96}, no.7, 076006 (2017)
%doi:10.1103/PhysRevD.96.076006
[arXiv:1709.07331 [hep-ph]].
%27 citations counted in INSPIRE as of 19 Dec 2022

%\cite{Kohlfurst:2022vwf}
\bibitem{Kohlfurst:2022vwf}
C.~Kohlf\"urst,
``The Heisenberg-Wigner formalism for transverse fields,''
[arXiv:2212.06057 [hep-ph]].
%0 citations counted in INSPIRE as of 19 Dec 2022

%\cite{Brezin:1970}
\bibitem{Brezin:1970}
E.~Brezin and C.~Itzykson,
``Pair Production in Vacuum by an Alternating Field,''
Phys. Rev. D 2, 1191 (1970)

%\cite{Popov:1972}
\bibitem{Popov:1972}
V.~S.~Popov,
``Pair Production in a Variable and Homogeneous Electric Field as an Oscillator Problem'',
JETP \textbf{35} 659 (1972)

%\cite{Popov:2005}
\bibitem{Popov:2005}
V.~S.~Popov,
``Imaginary-time method in quantum mechanics and field theory,''
Phys. Atom. Nucl. \textbf{68}, 686 (2005)

%\cite{Dunne:2005wi}
\bibitem{Dunne:2005wi}
G.~V.~Dunne and C.~Schubert, 
``Worldline instantons and pair production in inhomogeneous fields,'' 
Phys. Rev. D \textbf{72}, 105004 (2005) 
[arXiv:hep-th/0507174 [hep-th]].

%\cite{Dunne:2006fp}
\bibitem{Dunne:2006fp}
G.~V.~Dunne, Q.~h.~Wang, H.~Gies and C.~Schubert,
``Worldline instantons. II. The Fluctuation prefactor,''
Phys. Rev. D \textbf{73}, 065028 (2006)
[arXiv:hep-th/0602176[hep-th]]

%\cite{Kohlfurst:2021skr}
\bibitem{Kohlfurst:2021skr}
C.~Kohlf\"urst, N.~Ahmadiniaz, J.~Oertel and R.~Sch\"utzhold,
``Sauter-Schwinger effect for colliding laser pulses,''
[arXiv:2107.08741 [hep-ph]].
%6 citations counted in INSPIRE as of 25 Nov 2022

%\cite{Affleck:1981bma}
\bibitem{Affleck:1981bma}
I.~K.~Affleck, O.~Alvarez and N.~S.~Manton,
``Pair Production at Strong Coupling in Weak External Fields,''
Nucl. Phys. B \textbf{197}, 509-519 (1982)
%doi:10.1016/0550-3213(82)90455-2
%208 citations counted in INSPIRE as of 05 Dec 2021

%\cite{Dunne:2006ur}
\bibitem{Dunne:2006ur}
G.~V.~Dunne and Q.~h.~Wang,
``Multidimensional Worldline Instantons,''
Phys. Rev. D \textbf{74}, 065015 (2006)
%doi:10.1103/PhysRevD.74.065015
[arXiv:hep-th/0608020 [hep-th]].
%64 citations counted in INSPIRE as of 25 Nov 2022

%\cite{Dumlu:2011cc}
\bibitem{Dumlu:2011cc}
C.~K.~Dumlu and G.~V.~Dunne,
``Complex Worldline Instantons and Quantum Interference in Vacuum Pair Production,''
Phys. Rev. D \textbf{84}, 125023 (2011)
%doi:10.1103/PhysRevD.84.125023
[arXiv:1110.1657 [hep-th]].
%76 citations counted in INSPIRE as of 21 Dec 2022

%\cite{Gould:2017fve}
\bibitem{Gould:2017fve}
O.~Gould and A.~Rajantie,
``Thermal Schwinger pair production at arbitrary coupling,''
Phys. Rev. D \textbf{96}, no.7, 076002 (2017)
%doi:10.1103/PhysRevD.96.076002
[arXiv:1704.04801 [hep-th]].
%42 citations counted in INSPIRE as of 16 Jul 2021

%\cite{Torgrimsson:2017cyb}
\bibitem{Torgrimsson:2017cyb}
G.~Torgrimsson, C.~Schneider and R.~Sch\"utzhold,
``Sauter-Schwinger pair creation dynamically assisted by a plane wave,''
Phys. Rev. D \textbf{97}, no.9, 096004 (2018)
%doi:10.1103/PhysRevD.97.096004
[arXiv:1712.08613 [hep-ph]].
%29 citations counted in INSPIRE as of 31 Oct 2022

%\cite{Schneider:2018huk}
\bibitem{Schneider:2018huk}
C.~Schneider, G.~Torgrimsson and R.~Sch\"utzhold,
``Discrete worldline instantons,''
Phys. Rev. D \textbf{98}, no.8, 085009 (2018)
%doi:10.1103/PhysRevD.98.085009
[arXiv:1806.00943 [hep-th]].
%17 citations counted in INSPIRE as of 31 Oct 2022

%\cite{Gonoskov:2012}
\bibitem{Gonoskov:2012}
I.~Gonoskov, A.~Aiello, S.~Heugel, and G.~Leuchs,
``Dipole pulse theory: Maximizing the field amplitude from $4\pi$ focused laser pulses,'' 
Physical Review A \textbf{86}, 053836 (2012).

%\cite{Gonoskov:2013ada}
\bibitem{Gonoskov:2013ada}
A.~Gonoskov, I.~Gonoskov, C.~Harvey, A.~Ilderton, A.~Kim, M.~Marklund, G.~Mourou and A.~M.~Sergeev,
``Probing nonperturbative QED with optimally focused laser pulses,''
Phys. Rev. Lett. \textbf{111}, 060404 (2013)
%doi:10.1103/PhysRevLett.111.060404
[arXiv:1302.4653 [hep-ph]].
%53 citations counted in INSPIRE as of 06 Dec 2021

%\cite{Feynman:1948}
\bibitem{Feynman:1948}
R.~P.~Feynman,
``Space-time approach to nonrelativistic quantum mechanics'',
Rev. Mod. Phys. \textbf{20}, 367-387 (1948)

%\cite{Feynman:1950}
\bibitem{Feynman:1950}
R.~P.~Feynman,
``Mathematical formulation of the quantum theory of electromagnetic interaction,''
Phys. Rev. \textbf{80}, 440-457 (1950)

%\cite{Feynman:1951gn}
\bibitem{Feynman:1951gn}
R.~P.~Feynman,
``An Operator calculus having applications in quantum electrodynamics,''
Phys. Rev. \textbf{84}, 108-128 (1951)
%doi:10.1103/PhysRev.84.108
%419 citations counted in INSPIRE as of 08 Dec 2021

%\cite{DegliEsposti:2021its}
\bibitem{DegliEsposti:2021its}
G.~Degli Esposti and G.~Torgrimsson,
``Worldline instantons for nonlinear Breit-Wheeler pair production and Compton scattering,''
Phys. Rev. D \textbf{105}, no.9, 096036 (2022)
%doi:10.1103/PhysRevD.105.096036
[arXiv:2112.11433 [hep-ph]].
%3 citations counted in INSPIRE as of 22 Aug 2022

%\cite{Barut:1989mc}
\bibitem{Barut:1989mc}
A.~O.~Barut and I.~H.~Duru,
``Pair Production In An Electric Field In A Time Dependent Gauge,''
Phys.\ Rev.\ D {\bf 41} (1990) 1312.
%doi:10.1103/PhysRevD.41.1312
%%CITATION = doi:10.1103/PhysRevD.41.1312;%%
%2 citations counted in INSPIRE as of 12 Jan 2018  

%\cite{Rajeev:2021zae}
\bibitem{Rajeev:2021zae}
K.~Rajeev,
``Lorentzian worldline path integral approach to Schwinger effect,''
Phys. Rev. D \textbf{104}, no.10, 105014 (2021)
%doi:10.1103/PhysRevD.104.105014
[arXiv:2105.12194 [hep-th]].
%2 citations counted in INSPIRE as of 23 Nov 2021

%\cite{Strassler:1992zr}
\bibitem{Strassler:1992zr}
M.~J.~Strassler,
``Field theory without Feynman diagrams: One loop effective actions,''
Nucl. Phys. B \textbf{385}, 145-184 (1992)
%doi:10.1016/0550-3213(92)90098-V
[arXiv:hep-ph/9205205 [hep-ph]].
%451 citations counted in INSPIRE as of 16 Dec 2022

%\cite{Schubert:2001}
\bibitem{Schubert:2001}
C.~Schubert,
``Perturbative quantum field theory in the string-inspired formalism,'' Phys. Rept. \textbf{355}, 73 (2001) [arXiv:hep-th/0101036].

%\cite{Fradkin:1991ci}
\bibitem{Fradkin:1991ci}
E.~S.~Fradkin and D.~M.~Gitman,
``Path integral representation for the relativistic particle propagators and BFV quantization,''
Phys. Rev. D \textbf{44}, 3230-3236 (1991)
%doi:10.1103/PhysRevD.44.3230
%115 citations counted in INSPIRE as of 08 Dec 2021

%\cite{Corradini:2020prz}
\bibitem{Corradini:2020prz}
O.~Corradini and G.~Degli~Esposti,
``Dressed Dirac propagator from a locally supersymmetric $N=1$ spinning particle,''
Nucl. Phys. B \textbf{970}, 115498 (2021)
%doi:10.1016/j.nuclphysb.2021.115498
[arXiv:2008.03114 [hep-th]].
%3 citations counted in INSPIRE as of 08 Dec 2021

%\cite{Gies:2005ke}
\bibitem{Gies:2005ke}
H.~Gies and J.~Hammerling,
``Geometry of spin-field coupling on the worldline,''
Phys. Rev. D \textbf{72}, 065018 (2005)
%doi:10.1103/PhysRevD.72.065018
[arXiv:hep-th/0505072 [hep-th]].
%5 citations counted in INSPIRE as of 08 Dec 2021

%\cite{Ahmadiniaz:2020wlm}
\bibitem{Ahmadiniaz:2020wlm}
N.~Ahmadiniaz, V.~M.~Banda Guzm\'an, F.~Bastianelli, O.~Corradini, J.~P.~Edwards and C.~Schubert,
``Worldline master formulas for the dressed electron propagator. Part I. Off-shell amplitudes,''
JHEP \textbf{08}, no.08, 049 (2020)
%doi:10.1007/JHEP08(2020)018
[arXiv:2004.01391 [hep-th]].
%8 citations counted in INSPIRE as of 30 Jul 2021


\end{thebibliography}
\end{document}